\documentclass[12pt,twoside]{article}
\usepackage{graphicx,amsmath,amssymb,latexsym}
\usepackage{setspace}
\newcommand{\SM}{$SU(3)\times SU(2)_L\times U(1)_Y$ }

\DeclareMathOperator{\Tr}{Tr} \DeclareMathOperator{\tr}{tr}

\newcommand{\be}{\begin{equation}} \newcommand{\ee}{\end{equation}}
\newcommand{\barr}{\left(\begin{array}}
\newcommand{\earr}{\end{array}\right)}

\newcommand{\Yo}[1]{\kappa_{#1}}
\newcommand{\fr}[2]{\frac{#1}{#2}}
\newcommand{\Ys}{Y_2(S)}
\newcommand{\thh}{\frac{3}{2}}

\newcommand{\nf}{\frac{9}{4}}
\newcommand{\betfn}[1]{(4 \pi)^2 \frac{d #1}{d t}}
\newcommand{\la}{\lambda}

\begin{document}
\pagestyle{myheadings} \markboth{Is Dark Matter Heavy Because of Electroweak Symmetry Breaking?}{Revisiting Heavy Neutrinos}

\title{{\bf Is Dark Matter Heavy Because of Electroweak Symmetry Breaking? Revisiting Heavy Neutrinos}}
\author{Philip C. Schuster\footnote{Email:
 schuster@fas.harvard.edu}\hspace{.1cm} and Natalia
 Toro\footnote{Email: toro@fas.harvard.edu}\\ \\
{\it Jefferson Laboratory of Physics, Harvard University} \\ {\it Cambridge, Massachusetts 02138, USA} }
\date{\small\today} \maketitle
\begin{abstract}
\noindent A simple and well-motivated explanation for the origin of dark matter is that it consists of thermal relic particles that get their
mass entirely through electroweak symmetry breaking. The simplest models implementing this possibility predict a dark matter candidate that
consists of a mixture of two Dirac neutrinos with opposite isospin, and so has suppressed coupling to the $Z$. These models predict dark matter
masses of $m_{DM}\approx 45$ GeV or $m_{DM}\approx 90-95$ GeV and WIMP-neutron spin-independent cross sections $\sigma_{WIMP-n}\sim
10^{-6}-10^{-8}$ pb. Current direct dark matter searches are probing a portion of the parameter space of these models while future experiments
sensitive to $\sigma_{WIMP-n}\sim 10^{-8}$ pb will probe the remainder. An enhancement of the galactic halo gamma ray and positron flux coming
from annihilations of these particles is also expected across the $\sim 1-100$ GeV range. The framework further suggests an environmental
explanation of the hierarchy between the weak and Planck scales and of the small value of the cosmological constant relative to the weak scale.
\end{abstract}

\newpage
\section{Introduction}\label{sec:intro}
Stable particles produced thermally in the early universe with weak interactions and mass near $\sim 500$ GeV can account for the measured
density of dark matter. This suggests a possible relationship between dark matter and electroweak physics. Consequently, most efforts to explain
the origin of dark matter have focused on the Lightest Super-Partner (LSP) in the Minimal Supersymmetric Standard Model (MSSM) \cite{MSSM}. This
possibility is attractive because the presence of weak-scale supersymmetry may resolve the gauge hierarchy problem, predicts gauge coupling
unification, and explains the connection between the dark matter scale and weak scale. Yet weak-scale supersymmetry was not discovered at LEPII
nor have any conclusive indirect signatures been seen \cite{Eidelman:2004wy}. In light of this, the possibility that nature is seemingly
fine-tuned has received renewed interest \cite{Weinberg:1987dv,Vilenkin:1994ua}, and so we reconsider a naive explanation for the existence of
weak-scale dark matter that does not rely on supersymmetry: {\em Can it be that the dark matter scale coincides with the electroweak scale
because the dark matter particles get their mass entirely from electro-weak symmetry breaking?}

In this paper, we show that the answer to this question is yes, that such a framework is consistent with experimental data, and that it does not
require elaborate fine-tunings. Moreover, it is surprisingly predictive and suggests resolutions to both the gauge and cosmological constant
hierarchy problems within a landscape framework.

In Section \ref{sec:darkmatter} we review the evidence for dark matter and briefly consider the experimental constraints on heavy ($\sim 20-100$
GeV) neutrinos produced thermally as dark matter candidates. We conclude that, unlike neutrinos with invariant mass terms, a single neutrino
with only a Higgs-Yukawa mass is always under-abundant. We note, however, that suppression of the neutrino coupling to the $Z$ boson can remedy
the many problems with a single heavy neutrino.

In Section \ref{sec:EWSBdarkmatter} we study the effects of $Z$ suppression on the thermal heavy neutrino density. The suppression can be
realized by adding to the Standard Model with a Higgs doublet a vector-like ``dark sector'' consisting of two lepton generations with opposite
hypercharge and two neutral singlets (\ref{subsec:toy}). A chiral symmetry prevents these states from mixing with standard model particles or
acquiring invariant mass terms. After electroweak symmetry breaking, the stable $U(1)_{EM}$-neutral particle is a mixture of opposite-isospin
neutral states, thereby suppressing its coupling to the $Z$. We present calculations of the dark matter abundance in this toy model
(\ref{subsec:relicdensity}).

In Section \ref{sec:toys}, we revisit the model of \ref{subsec:toy} and another two-neutrino model allowing $Z$-coupling suppression, focusing
on the radiative stability of the near-maximal mixing required to reproduce the measured $\Omega_{DM}$. The second model
(\ref{subsec:toytriplet}) is reminiscent of the particle content of Split Supersymmetry\cite{Arkani-Hamed:2004fb}, with the gluino removed and
supersymmetric relations badly broken. We also point out in (\ref{sec:unification}) that gauge coupling unification can be achieved at $M_G\sim
10^{11}$ GeV by extending the dark sector to include three of the vector lepton families of Sec. \ref{subsec:toy}.

In Section \ref{sec:Experiment}, we consider the consistency of the toy models of Sections \ref{sec:EWSBdarkmatter} and \ref{sec:toys} with
current experimental constraints, and the prospects for detection in the near future in direct (\ref{subsec:directdm}) and indirect
(\ref{subsec:indirectdm}) dark matter searches, as well as in colliders (\ref{subsec:collider}). We focus on the region of parameter space that
is most consistent with the observed dark matter density. We also consider the constraints placed on the model by precision electroweak
observables (\ref{subsec:pew}).

Although our motivation is independent of the weak-Planck hierarchy problem, the models we consider do suggest possible resolutions to this
problem within a landscape framework. The crucial link in this argument is Weinberg's observation that structure formation requires that the
cosmological constant be small \cite{Weinberg:1987dv}. By explicitly connecting the dark matter mass to the Higgs vev, the class of models
suggested here relates the weak-$\Lambda$ hierarchy to the weak-Planck hierarchy. We also further develop the suggestion in
\cite{Arkani-Hamed:2005yv} that the gauge hierarchy may be explained using the structure principle because the existence of massive dark matter
is tied to the stability of the Higgs vacuum. These two arguments are discussed in Section \ref{sec:structure}.

\section{Dark Matter: Evidence and Constraints}\label{sec:darkmatter}
The existence of dark matter is by now well established. The most direct evidence comes from comparing the luminous mass of a galaxy or cluster
to its total mass, inferred by indirect means. Measurements of galactic rotation curves indicate that the mass of galaxies must be greater and
more spread out than the luminous matter they contain, suggesting a dark matter component $\Omega_{DM} \geq 0.1$. Various means of estimating
the mass of galaxy clusters, such as velocity dispersions and gravitational lensing, imply $\Omega_{DM} \approx 0.2-0.3$ on the cluster scale. A
less direct---but more precise---probe of dark matter is the anisotropy of the cosmic microwave background (CMB)\cite{Spergel:2003cb}. Assuming
a $\Lambda$CDM model (in which all unknown energy density is assumed to come from either a cosmological constant or non-relativistic dark
matter), current measurements indicate a total matter component of $\Omega_M h^2=0.134 \pm 0.006$, with a much smaller baryonic matter component
of $\Omega_B h^2= 0.023\pm 0.001$ (which is consistent with the limits on $\Omega_B$ from the abundances of light elements produced in big-bang
nucleosynthesis\cite{Eidelman:2004wy}). Along with MACHO searches\cite{Alcock:2000ph,Afonso:2002xq}, this is the principal observational
evidence for the non-baryonic nature of at least some of the dark matter. Finally, detailed modeling of our own galaxy suggest a local dark
matter density of $\rho_{DM}\approx 0.3$ GeV/$cm^3$. For more careful discussions of the evidence for dark matter and its properties, we refer
the reader to the reviews \cite{Bertone:2004pz,Gondolo:2004fg,Griest:1989wd}.

In light of the increasingly precise evidence for a sizable dark matter component, many explanations of its nature and origin have been proposed
(see \cite{Bertone:2004pz,Gondolo:2004fg,Griest:1989wd} for a review of both particle physics-motivated and purely astrophysical candidates).
The elementary particle candidates are either (1) thermal relics, which begin in thermal equilibrium in the early universe but decouple from the
plasma as the expansion rate of the universe exceeds their annihilation rates or (2) non-thermal relics, produced out of equilibrium  (for
example, from the decay of another particle that has itself decoupled from the plasma). The models we propose are of the first class. We further
specialize to the case of dark matter particles that become non-relativistic before they decouple.

In the hot early universe, the dark matter particles are kept in chemical and kinetic equilibrium with the surrounding particle bath by pair
production, annihilation, and scattering. If their interactions with the bath are sufficiently strong, they remain coupled until after they
become non-relativistic. As the universe cools below $T \approx m_{DM}$, their density becomes exponentially Boltzmann-suppressed, quenching the
annihilations. This ``freeze-out'' happens quite suddenly, typically at a temperature of $T_f\approx \frac{m_{DM}}{25}$. From this point on, the
co-moving particle density remains approximately constant.  Thus, we can compare the observed dark matter density today to the predicted
abundance, which is very roughly
\begin{equation}
\Omega_{DM}\approx \frac{0.1 pb \times c}{\langle \sigma v\rangle}. \label{dmcomponentcalc}
\end{equation}

Assuming the maximum perturbative cross section $\sim \frac{1}{m_{DM}^2}$ for the dark matter particle's annihilations and requiring consistency
of Equation (\ref{dmcomponentcalc}) with the observed dark matter component places an upper bound on its mass, $m_{DM} \lesssim$ 300 TeV
\cite{Griest:1989wd}. For a weak-interaction coupling, the mass scale of interest is near a TeV, coincidentally close to the scale of
electroweak symmetry breaking.  Weakly interacting particles with mass near a TeV (WIMPs) provide especially intriguing dark matter candidates
because of their possible connection to electroweak physics.

One simple WIMP candidate is a stable heavy Dirac or Majorana neutrino. Neutrinos with a Dirac- or Majorana-type invariant mass have small
enough annihilation rates to account for the observed dark matter abundance if they are heavier than $\approx 500$ GeV, though this option seems
to be ruled out by direct dark matter searches (see Sec. \ref{subsec:directdm}). The best-motivated and currently well-studied WIMP candidate is
the LSP in supersymmetric extensions of the Standard Model. In these models, the dark matter mass and electroweak symmetry breaking scale are
close for a reason---because both are determined by the scale of supersymmetry breaking.

Another possible explanation of this proximity of scales is if, whatever may set the EWSB scale, the symmetry breaking itself sets the scale of
the dark matter mass, which comes entirely from a Higgs Yukawa coupling. Neutrinos with Standard Model coupling to the $Z$ that get mass from
electroweak symmetry breaking are well ruled out by experiments. $Z$-pole width measurements at LEP have ruled out neutrinos lighter than
$\frac{M_{Z}}{2}$, and a mass above $\frac{M_{Z}}{2}$ would lead to over-annihilation.  Thus $\Omega_{DM} h^2=0.11\pm.011$ cannot be obtained
from a Standard Model-like neutrino of any mass. We revisit the possibility that the dark matter particle is given mass by electroweak symmetry
breaking and observe that the problems associated with heavy neutrino dark matter can be resolved if the particle's coupling to the $Z$ is
suppressed.

\section{Dark Matter with Suppressed $Z$-Coupling}\label{sec:EWSBdarkmatter}
\subsection{A Toy Model} \label{subsec:toy}
We consider a model suggested in \cite{Arkani-Hamed:2005yv} consisting of the minimal Standard Model, a lepton generation ($L_1$, $\bar E_1$)
with the usual standard model charges, and one ($L_2$,$\bar E_2$) with opposite hypercharge assignments. To this we add two \SM singlets ($s_1$,
$s_2$) (this particle content is among the smallest anomaly-free extensions of the Standard Model in which all particles can obtain mass via EW
symmetry breaking). A chiral symmetry under which the Standard Model is neutral and the two new lepton doublets have charge opposite that of the
four new singlets (or a discrete subgroup thereof)  assures that all invariant mass terms vanish and mixing with Standard Model leptons is
eliminated, making the lightest particle exactly stable. This chiral symmetry plays a role analogous to that played by R-parity in
supersymmetric models.

The most general Lagrangian consistent with this symmetry takes the form
\begin{equation}
\mathcal{L}_{\hbox{int}} = Y_1 h^c L_1 \bar E_1 + Y_2 h L_2 \bar E_2 + (h L_1, h^c L_2) K \left(\begin{array}{c} s_1 \\ s_2
\end{array} \right),
\end{equation}
where $h^c=- i \sigma_2 h^*$, and implicit products of $SU(2)$ doublets are taken to mean $AB \equiv A^T i \sigma_2 B$. Up to re-phasing, the
$2\times 2$ matrix $K$ can be written as
\begin{equation}
K=\barr{cc} \cos \theta & \sin \theta \\
-\sin \theta & \cos \theta \earr \barr{cc} \kappa & 0 \\ 0 & \kappa' \earr \mathcal{V} ,
\end{equation}
with $\mathcal{V}$ unitary.

In terms of mass eigenstates, the Higgs- and $Z$-interaction Lagrangian for the $U(1)_{EM}$-neutral states takes the form
\begin{equation}
\mathcal{L} \supset (1+\frac h v)(m N s+m' N' s') +\frac g {2 \cos\theta_W} Z_\mu (\bar N,\bar N') \bar \sigma^\mu \barr{cc} \cos(2 \theta) &
\sin(2 \theta) \\ \sin(2 \theta) & \cos(2 \theta) \earr \barr{c} s \\ s' \earr,
\end{equation}
where
\begin{equation} \binom{N}{N'}=\binom{N_1 \cos(\theta) -N_2 \sin(\theta)}{N_1 \sin(\theta) +N_2 \cos(\theta)},\quad
\binom{s}{s'}=\mathcal{V}\binom{s_1}{s_2},
\end{equation}
and $m=\frac v {\sqrt 2} \kappa$ and $m'=\frac v {\sqrt 2} \kappa'$ are the neutral particle masses after $h$ takes the vev
$\left<h\right>=\frac{1}{\sqrt{2}}\dbinom{0}{v}$. The strength of the neutral fermions' coupling to the $Z$ is suppressed by $\epsilon \equiv
\cos(2\theta)$ compared to that of a Standard Model neutrino. As we shall see, in the $\theta \approx \frac \pi 4$ regime, the particle is a
viable dark matter candidate.  Henceforth, we refer to the lighter neutral state as a Dirac fermion $\chi$, and the heavier as $\chi'$.

\subsection{Relic Density}\label{subsec:relicdensity}
\begin{figure}
\begin{center}
\includegraphics[width=12cm]{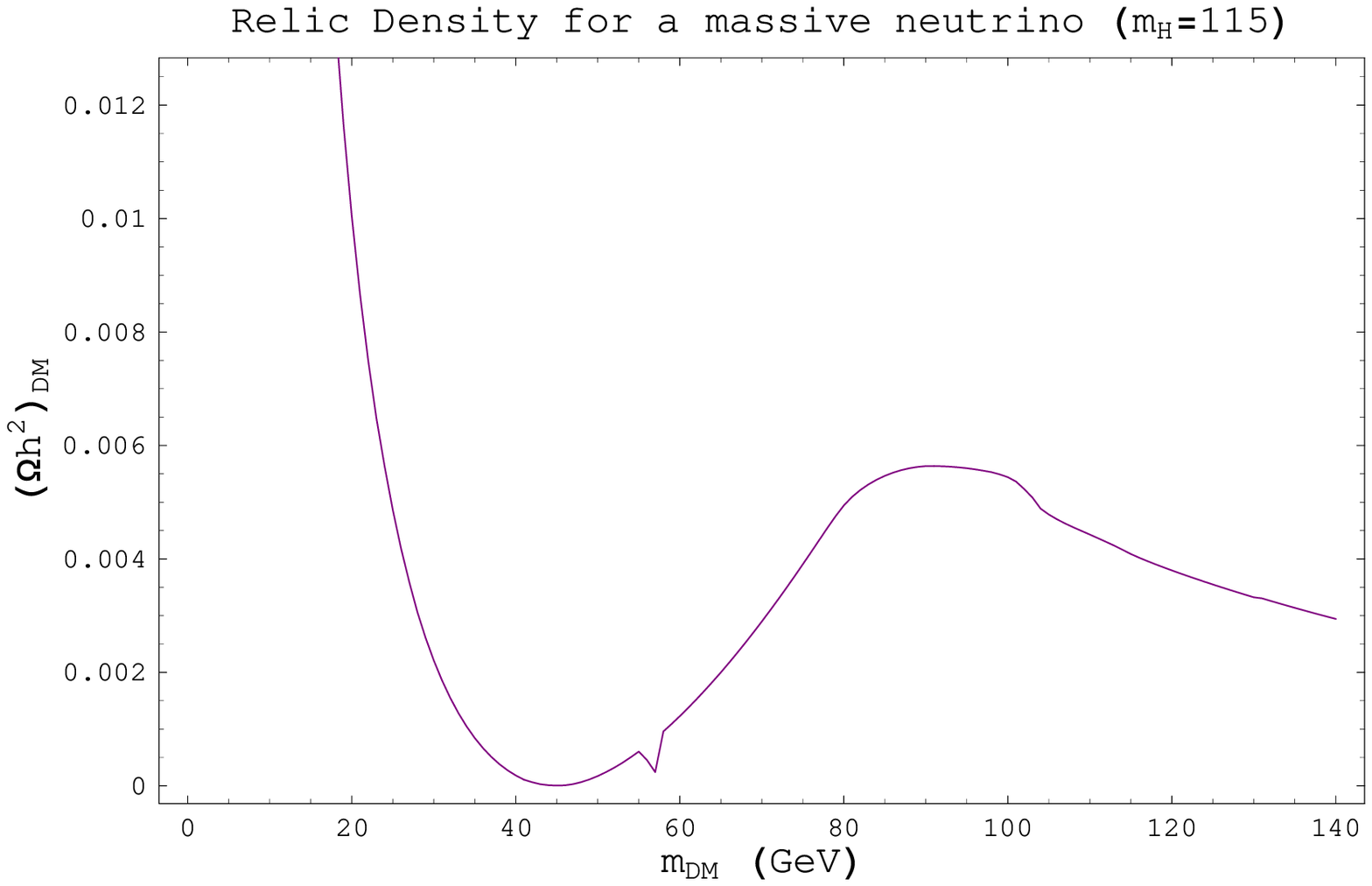}
\includegraphics[width=10cm,angle=0]{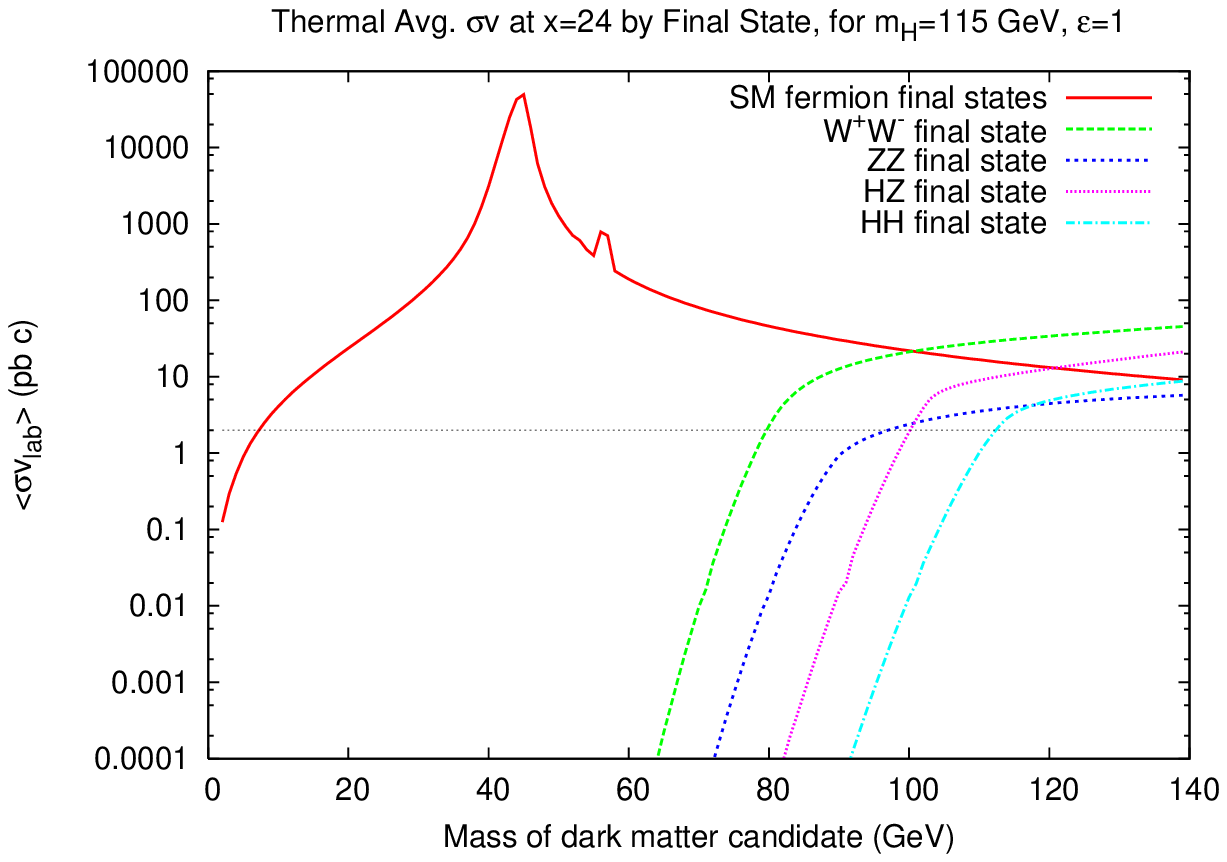}
\end{center}
\caption{Top: Relic density versus mass for a heavy Dirac Standard Model neutrino. Bottom: $\langle \sigma v \rangle$ at $x=24$ (near
freeze-out) versus mass for a Dirac Standard Model neutrino. Both plots assume a Higgs mass of $m_h=115$ GeV.} \label{fig:RelicDensitySM}
\end{figure}

The framework for calculating relic abundances for thermally produced, stable particles is discussed in Appendix \ref{app:relicdens}. For a
Standard Model Dirac neutrino, the relic density ignoring all co-annihilations is shown in Figure \ref{fig:RelicDensitySM} together with the
thermally averaged cross section $\langle \sigma v \rangle$ at $x=24 \approx x_f$. Reproducing the observed $\Omega_{DM}$ requires $\langle
\sigma v \rangle \approx 2$ pb c (double the estimate from \ref{dmcomponentcalc} because both $\chi$ and $\bar\chi$ contribute). As can be seen,
$Z$-mediated annihilations into fermion final states dominate below the $Z$-pole so that the relic density coincides with the experimentally
observed level only at $\approx 5$ GeV, a mass which has been ruled out by $Z$-width measurements at LEP. Even as the $Z$-mediated annihilation
rate drops with increasing $m_{DM}$, Higgs-mediated annihilations into $ZZ$ and $WW$ pairs are strong enough that the relic density never
exceeds $10^{-2}$. Thus, a neutrino with Higgs-Yukawa-type mass can only be the dark matter if the $Z$-coupling is suppressed.

The vector lepton toy model defined in \ref{subsec:toy} allows this suppression as does another model introduced in Sec.
\ref{subsec:toytriplet}. The $Z$-suppression in both cases results, as discussed above, from mass-mixing between gauge eigenstates with opposite
isospin.  There is always a corresponding enhancement of the $Z$-coupling between the heavy and light eigenstates, but if one is significantly
heavier than the other this has little effect on the relic density.

We have calculated the relic density of the lightest dark sector neutrino as a function of its mass $m_{DM}$ and $Z$-coupling suppression factor
$\epsilon=\cos{2\theta}$. Our calculations are based on the perturbative freezeout methods discussed in
\cite{Gondolo:1990dk,Griest:1990kh,Edsjo:1997bg} and the details of our calculation can be found in appendix \ref{app:relicdens}. We ignore the
effects of co-annihilations, which are unimportant because the dark matter state can be made naturally much lighter than the other states that
could co-annihilate with it.

\begin{figure}
\begin{center}
\includegraphics[width=12cm]{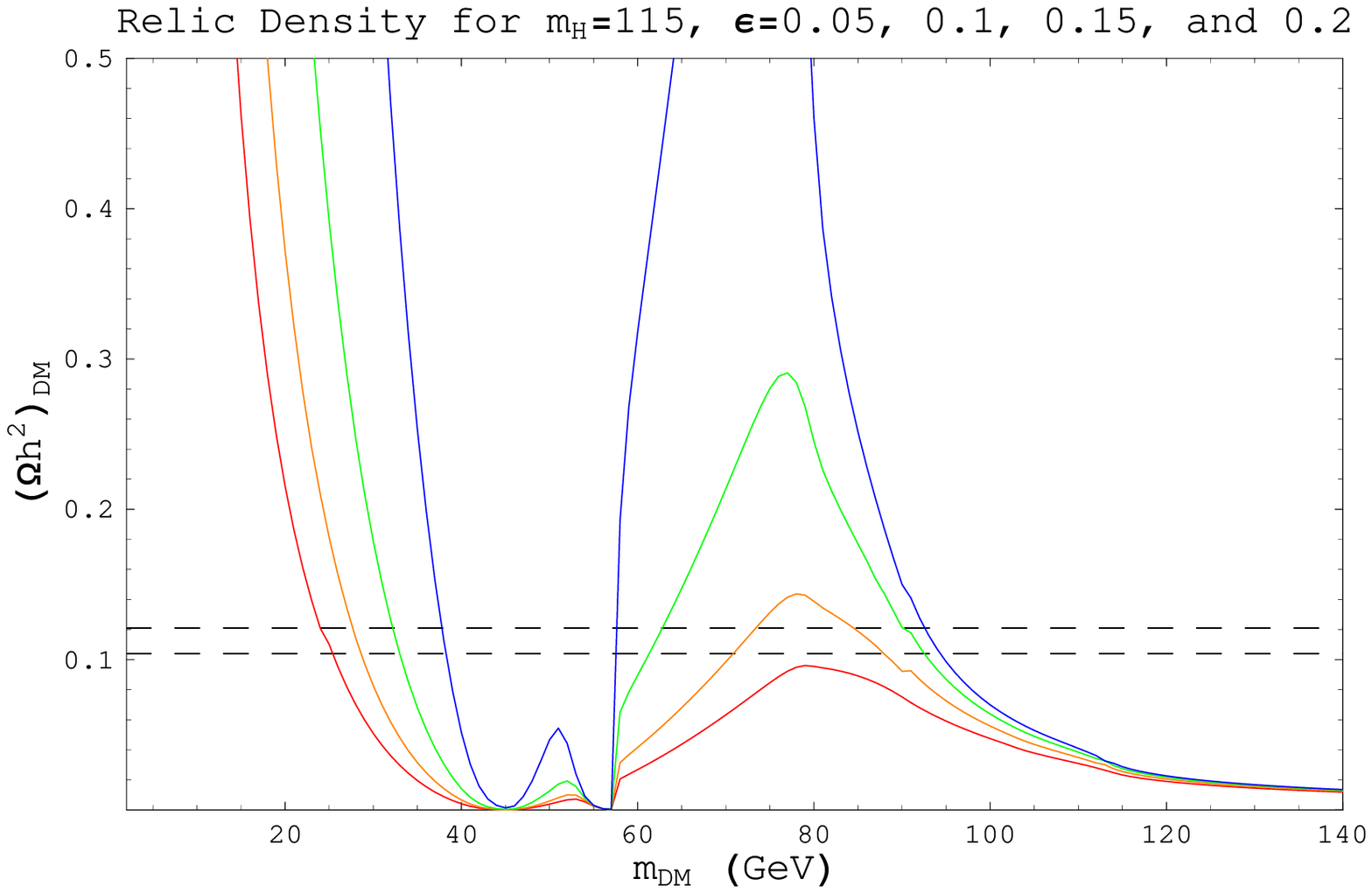}
\includegraphics[width=9cm,angle=0]{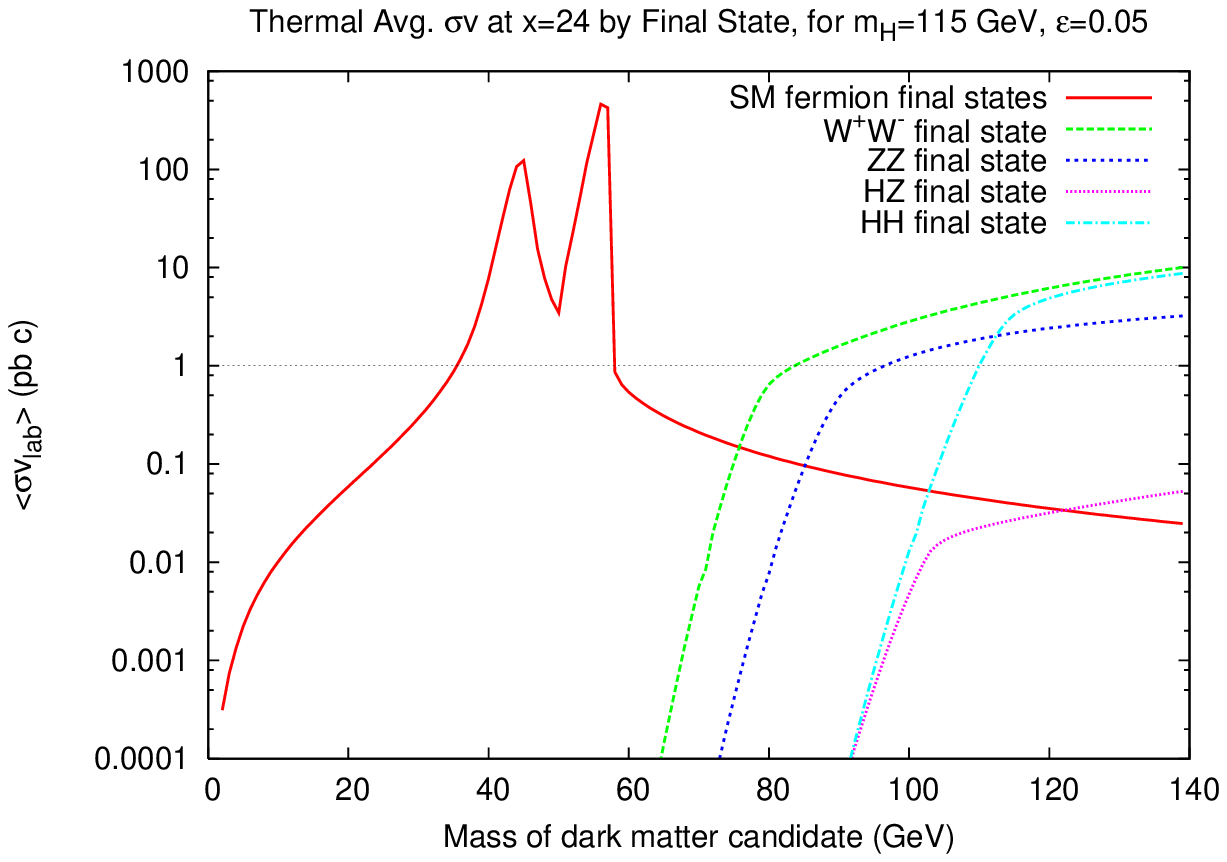}
\end{center}
\caption{Top: Relic density versus dark matter mass for $Z$ suppression of (top to bottom curves) 0.05, 0.1, 0.15, and 0.2. Bottom: Thermally
averaged annihilation cross section $\langle \sigma v \rangle$ by final state at $x=24$ (near freeze-out) versus mass, with $Z$ suppression of
0.05. Both plots assume a Higgs mass of $m_h=115$ GeV.} \label{fig:RelicDensity115}
\end{figure}

\begin{figure}
\begin{center}
\includegraphics[width=12cm]{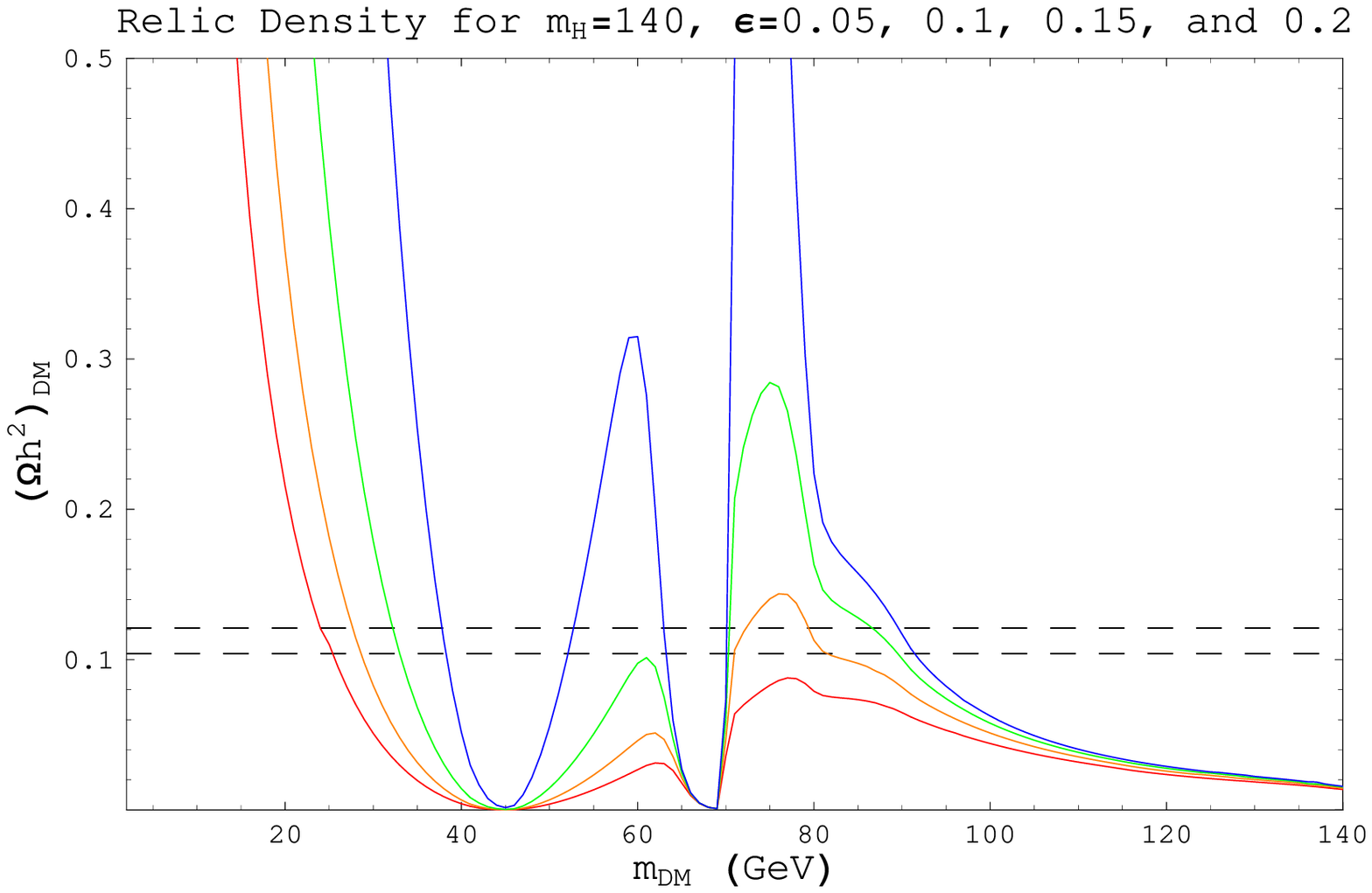}
\includegraphics[width=9cm,angle=0]{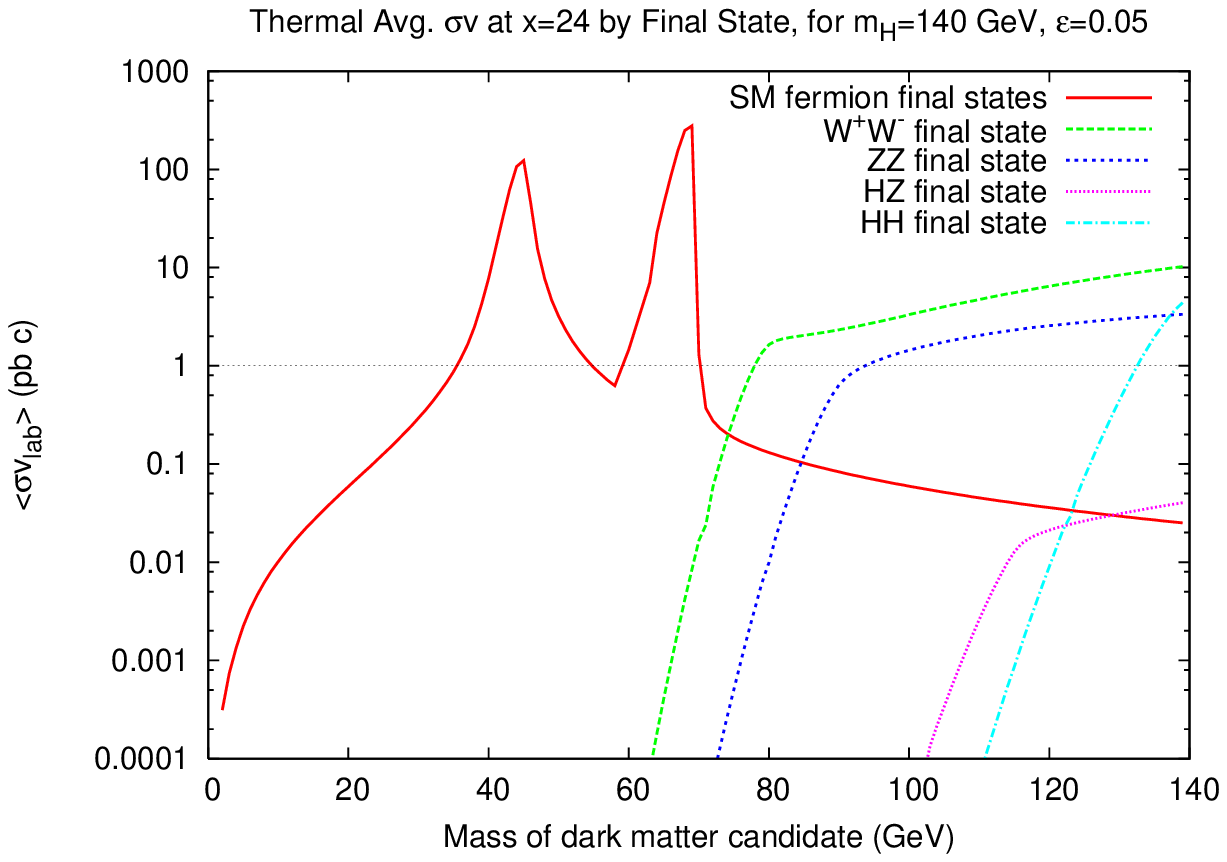}
\end{center}
\caption{Top: Relic density versus dark matter mass for $Z$ suppression of (top to bottom curves) 0.05, 0.1, 0.15, and 0.2. Bottom: Thermally
averaged annihilation cross section $\langle \sigma v \rangle$ by final state at $x=24$ (near freeze-out) versus mass, with $Z$ suppression of
0.05. Both plots assume a Higgs mass of $m_h=140$ GeV.} \label{fig:RelicDensity140}
\end{figure}

\begin{figure}
\begin{center}
\includegraphics[width=12cm]{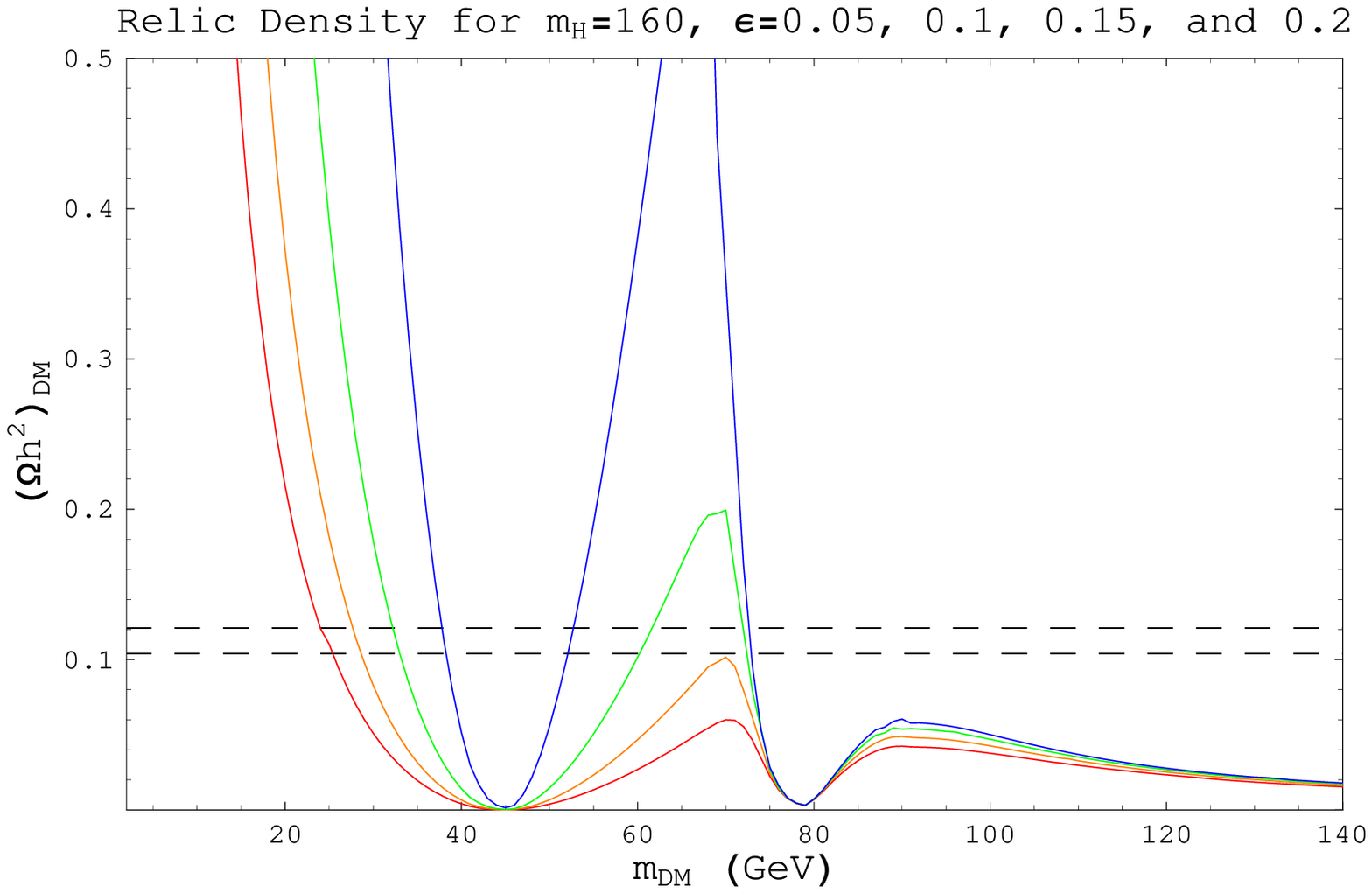}
\includegraphics[width=9cm,angle=0]{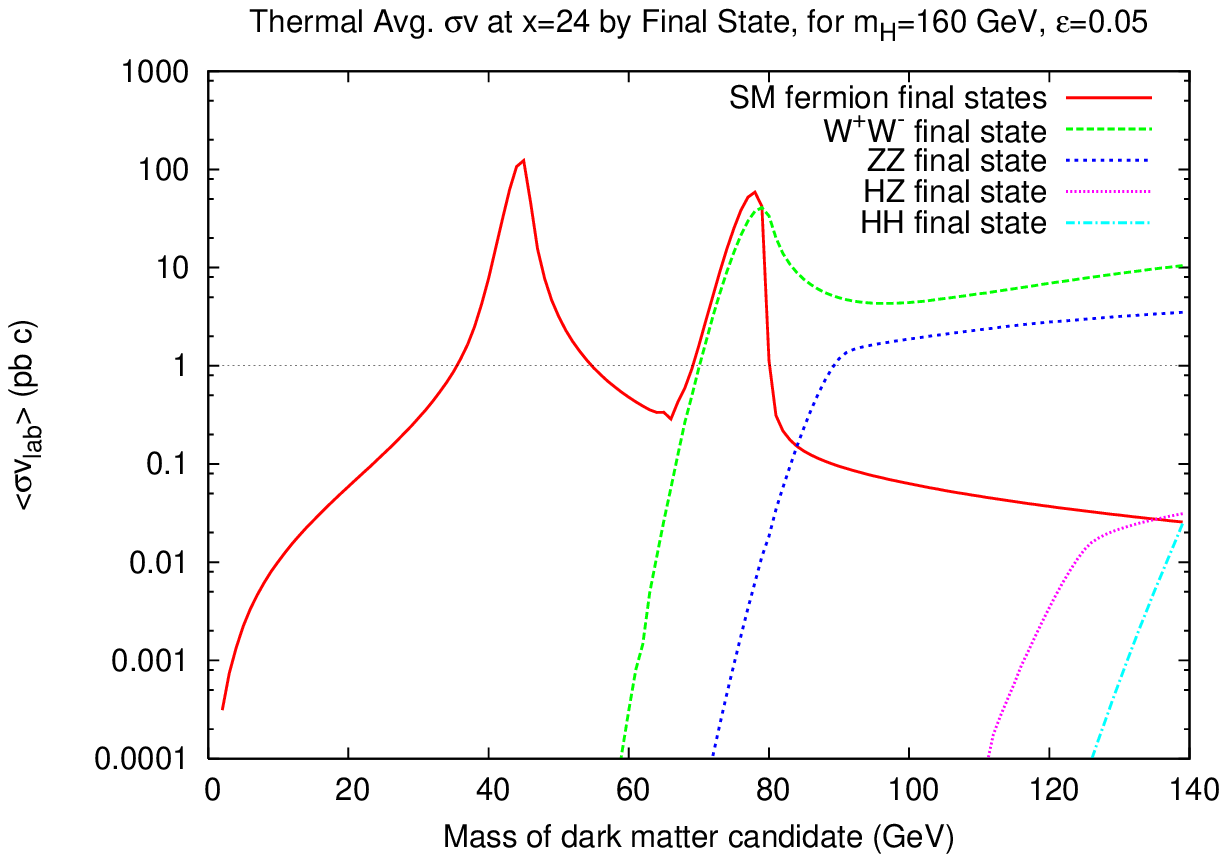}
\end{center}
\caption{Top: Relic density versus dark matter mass for $Z$ suppression of (top to bottom curves) 0.05, 0.1, 0.15, and 0.2. Bottom: Thermally
averaged annihilation cross section $\langle \sigma v \rangle$ by final state at $x=24$ (near freeze-out) versus mass, with $Z$ suppression of
0.05. Both plots assume a Higgs mass of $m_h=160$ GeV.} \label{fig:RelicDensity160}
\end{figure}

Figures \ref{fig:RelicDensity115}, \ref{fig:RelicDensity140}, and \ref{fig:RelicDensity160} display our results for Higgs masses of $m_h=115$,
140, and 160 GeV respectively. In each case, we consider $Z$-suppressions $\epsilon=0.05$, 0.1, 0.15, and 0.2. The highest curves correspond to
the lowest $\epsilon$, for which annihilation rates are lowest and the resulting relic densities highest.
The viable neutrino mass below the $Z$ pole is pushed up when the $Z$ coupling is suppressed, from $m_{DM} \approx 5$ GeV for a Standard Model
neutrino to $m_{DM}\approx 35$ GeV with $\epsilon\lesssim 0.1$. As we discuss in \ref{subsec:collider}, the $Z$-suppression of $0.1$ and
phase-space suppression from proximity of $m_{DM}$ to $m_Z/2$ is enough for a heavy neutrino in this mass range to have avoided detection at
LEP, while further suppression is required to evade the null results of many direct WIMP searches.

In the region $m_{DM}\geq \frac{M_Z}{2}$, there is a band of masses for which some $\epsilon \approx 0.1$ would produce the observed abundance.
This band is bounded by the $Z$-pole and by the onset of annihilation into $W^+ W^-$ and $ZZ$ pairs. This region sits near the $s$-channel Higgs
pole, so it is also chopped off where the efficient annihilation near the pole suppresses relic density. For a Higgs mass of $m_h=115$ GeV, the
relic density $\Omega h^2_{DM}$ is consistent with experiment in the range of $m_{DM}\approx 60-85$ GeV. On the $Z$-pole and above this region,
near $m_{DM}\approx 90$ GeV, the observed abundance can be reproduced by assuming very near-maximal mixing ($\epsilon \lesssim 0.01$).

When $Z$-suppression is allowed, there is an interesting region of parameter space in which a heavy neutrino can reproduce the $\Omega_{DM}$
inferred from observations. The supression necessary is, however, far greater than one would expect from a generic choice of Yukawa couplings.
In the following section, we examine the radiative stability of this minimal mixing and its possible origins.

\section{Models With Maximally Mixed Neutrinos}\label{sec:toys}
In this section, we discuss two minimal extensions of the Standard Model in which maximally mixed neutrinos with no
invariant mass accounts for the dark matter. The first is the toy model discussed above. The second has a particle
content reminiscent of Split-SUSY but with no supersymmetric relations among parameters.

\subsection{Vector Lepton Model}\label{subsec:toyvector}
As discussed above, the model of Section \ref{subsec:toy} generically mixes neutrino species of opposite isospin, resulting in supperssion of
the $Z$-coupling. But near-maximal mixing ($\epsilon \lesssim 0.15$) is necessary to reproduce the observed relic density and to evade exclusion
based on current data. We wish to consider the radiative stability of such small $\epsilon$.

The discrete symmetry under which
\begin{eqnarray}
(h,h^c) &\rightarrow & (h^c,-h) \\
(L_1,L_2) &\rightarrow & (-L_2,L_1) \\
(\bar{E_1},\bar{E_2}) &\rightarrow & (\bar{E_2},\bar{E_1}) \\
(s_1,s_2) &\rightarrow & (s_2,s_1)
\end{eqnarray}
implies that the neutral mass eigenstates are maximally mixed Dirac neutrinos and that the charged states are degenerate. It is, of course,
broken by Standard Model hypercharge and fermion mass differences. However, the Standard Model breaking does not generate a deviation from
maximum mixing at one-loop order. We have not analyzed the two-loop contributions carefully, but estimate roughly that, assuming this symmetry
is exactly preserved in the dark sector at $M_G=10^{16}$ GeV, the Standard Model breaking generates an $\epsilon \sim 0.001-0.01$ at the
electroweak scale. If the charged state Yukawa couplings are not exactly degenerate (but $\epsilon=0$ at $M_G$), then the non-degeneracy
$Y_1=Y_2 (1+\delta)$ at $M_G$ radiatively generates at the weak scale a $Z$ suppression
$$\epsilon \approx \frac{3}{16 \pi^2} \ln\left(\frac{M_G}{m_W}\right) \frac{K_1^2+K_2^2}{K_1^2-K_2^2} \delta,$$ where $K_1$ and $K_2$ are the
eigenvalues of $\kappa$. For reasonable values of the parameters, we find that $\epsilon \sim \delta$. These estimates are justified in Appendix
\ref{app:radgen}.

One can also consider an approximate $SU(2)$ symmetry that mixes the lepton doublets and rotates the Higgs. It acts as
\begin{eqnarray}
\mathcal{H} &=& (-h^c,h)\rightarrow \mathcal{U}_L \mathcal{H}
\mathcal{U}_R, \\
\mathcal{L} &=& (L_1,L_2)\rightarrow \mathcal{U}_L \mathcal{L} \mathcal{U}_R .
\end{eqnarray}
and can act on the Standard Model quark doublets as $\bar{Q}=(\bar{u},\bar{d})\rightarrow \bar{Q}\mathcal{U}_R$. The symmetry is thus an
extension of the ordinary Standard Model $SU(2)_R$ that enforces the custodial SU(2) $\rho=1$ relation, though it acts very differently on the
vector leptons than on the Standard Model fermions. The only invariant combination of $\mathcal{H}$ and $\mathcal{L}$ is
\begin{equation}
Tr[ (i \sigma_2^{(R)}) \mathcal{H}^T (i \sigma_2^{(L)}) \mathcal{L}] = h L_1 + h^c L_2
\end{equation}
Unlike the discrete symmetries above, it is broken by $Y_1$ and $Y_2$ even when they are degenerate, as well as by $K$ unless $K_{1i} = -
K_{2i}$ (i.e. $\theta=\pi/4$, with $\chi=\frac{1}{\sqrt{2}}(N_1+N_2)$ massless). In the limit that this symmetry is exact in the neutral sector,
the mass matrix takes the form
\begin{equation}
K=\frac{1}{\sqrt 2 } \barr{rr} 1 & 1 \\ -1 & 1 \earr \barr{cc} \kappa(\mu) & 0 \\ 0 & 0 \earr \barr{rr} \cos \alpha(\mu) & \sin \alpha(\mu)
\\ -\sin \alpha(\mu) & \cos \alpha(\mu) \earr. \label{eqn:symmetricdecomp}
\end{equation}
The condition $m_\chi=0$ is equivalent to demanding that some linear combination $s_0 \equiv -\sin \alpha s_1+\cos \alpha s_2$ of the singlets
is non-interacting, and so a mass for $\chi$ will never be generated radiatively within the \SM effective theory. Small explicit breaking is
needed and generically results in a $Z$-suppression $\epsilon$ of the same order as $m_\chi/m_{\chi'}$ where $m_{\chi'}$ is the mass of
$\chi'=\frac{1}{\sqrt{2}}(N_1-N_2)$.

\subsection{An Adjoint/Vector Lepton Model}\label{subsec:toytriplet}
Another minimal extension of the Standard Model that can naturally implement the electroweak dark matter scenario
consists of a vector pair of weak doublets $L_1$, $L_2$, an $SU(2)_L$ adjoint fermion $T=\barr{rr} T^0 & T^+ \\ T^- &
-T^0 \earr$, and a singlet $s$ (this is the particle content of Split-SUSY without a heavy gluino, and with no trace
of the supersymmetric relations). As in the vector lepton model, we impose a chiral symmetry to suppress the invariant
mass terms and mixing with the Standard Model. The Yukawa Lagrangian is
\begin{equation}
\mathcal{L}_{int}=Y_1 h T L_1 + Y_2 h^c T L_2+ Y_3 h L_1+ Y_4 h^c L_2.
\end{equation}
An approximate $SU(2)_R$ symmetry acting as
\begin{eqnarray}
\mathcal{H} &=& (-h^c,h)\rightarrow \mathcal{U}_L \mathcal{H}
\mathcal{U}_R \\
\mathcal{L} &=& (L_1,L_2)\rightarrow \mathcal{U}_L\mathcal{L}\mathcal{U}_R ,
\end{eqnarray}
with $T$ and $s$ both singlets under $SU(2)_R$ requires the Lagrangian to be of the form
\begin{eqnarray}
  \mathcal{L}_{int} &=&  Y^{\prime} Tr[ (i \sigma_2^{(R)}) \mathcal{H}^T (i \sigma_2^{(L)}) T \mathcal{L}]
                        +Y Tr[ (i \sigma_2^{(R)}) \mathcal{H}^T (i \sigma_2^{(L)}) \mathcal{L}]s \nonumber \\
   &=& Y^{\prime}(h T L_1 + h^c T L_2)+ Y s(h L_1+h^c L_2) ,
\end{eqnarray}
requiring $Y_1=Y_2=Y'$, $Y_3=Y_4=Y$. The corresponding low-energy mass matrix is
\begin{eqnarray}
\mathcal{L} &\supset & -\frac{v}{\sqrt 2} (N_1, N_2) \left(\begin{array}{cc}Y &Y^{\prime} \\ Y &
-Y^{\prime}\end{array}\right)
\left(\begin{array}{c}s \\
T^0 \end{array}\right) - \frac{Y^{\prime} v}{\sqrt{2}}(T^+E_1+T^-E_2) \\
&=& -(\chi, \chi^{\prime})\left(\begin{array}{cc}m & 0 \\ 0 &
m^{\prime}\end{array}\right) \left(\begin{array}{c}s \\
T^0 \end{array}\right) - \frac{m^{\prime}}{\sqrt{2}}(T^+E_1+T^-E_2) ,
\end{eqnarray}
where $m=Yv$, $m^{\prime}=Y^{\prime}v$, $\chi=\frac{1}{\sqrt{2}}(N_1+N_2)$, and $\chi^{\prime}=\frac{1}{\sqrt{2}}(N_1-N_2)$. Thus, $SU(2)_R$
generates maximally mixed neutrino mass eigenstates $\chi$ and $\chi^{\prime}$. Standard Model hypercharge and isodoublet mass splitting break
this symmetry, but communicate this breaking at two-loop. If all Yukawa terms are strongly coupled near the scale $M_G\approx 10^{16}$ GeV, then
we obtain a low energy spectrum near $\sim 100-140$ GeV. Because the tree-level charged state masses are smaller than the $\chi^{\prime}$ mass,
the dark matter must be the $\chi-s$ state. To reproduce $\Omega_{DM}$, we expect a $\chi-s$ mass in the range of $35-45$ GeV or $55-90$ GeV. We
have not studied the radiative effects in this model, but expect them to be of the same size as those discussed for the discrete symmetries of
the previous section.

\subsection{Gauge Coupling Unification}\label{sec:unification}
Both of the models discussed above can easily accommodate gauge unification, though this presents additional difficulties associated with proton
decay or large contributions to precision electroweak observables.

\begin{figure}
\begin{center}
\includegraphics[width=7cm]{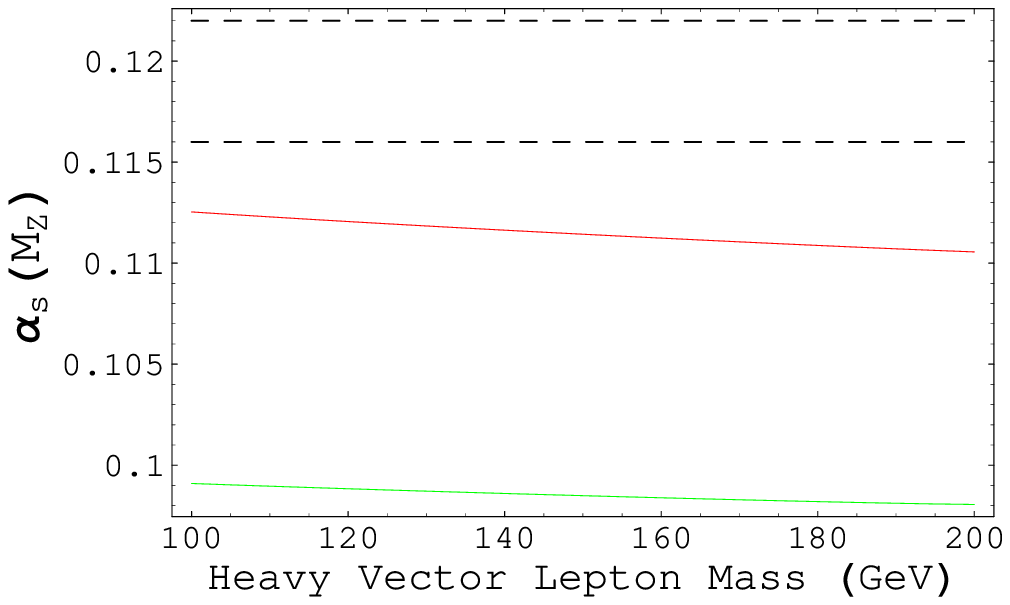}
\includegraphics[width=7cm]{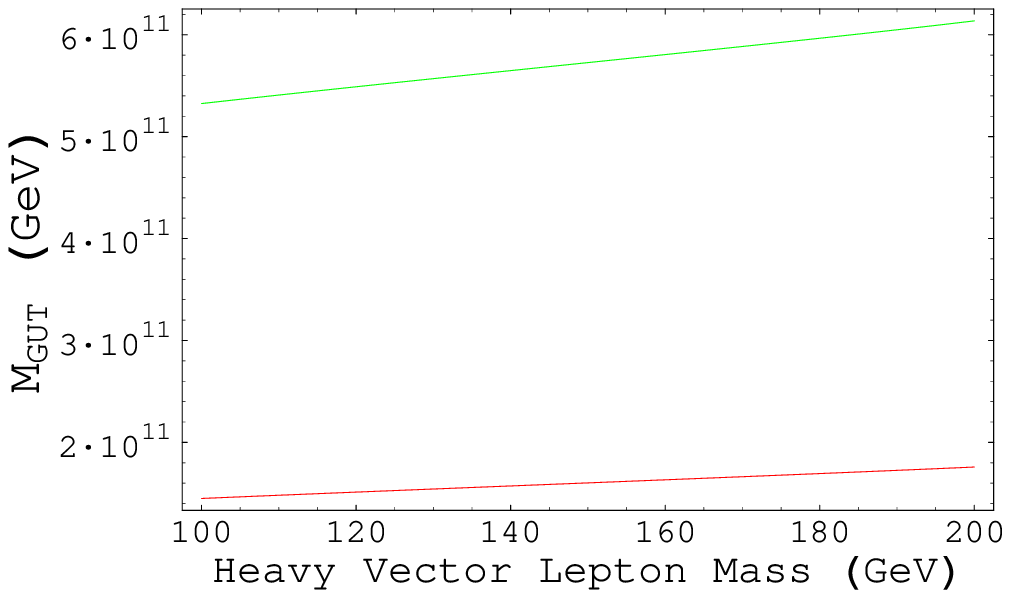}
\caption{Left: The top line (in red) shows the unification prediction for $\alpha_3(M_Z)$ for the three family vector lepton model while the
lower line (in green) shows the prediction for a two family model. The 1-$\sigma$ band for the experimentally measured $\alpha_3(M_Z)=0.119\pm
0.002$ \cite{Eidelman:2004wy} is shown by the dotted lines. Right: The top line (in green) shows the unification scale $M_G$ for a two family
vector lepton model while the bottom line (in red) shows $M_G$ for a three family model.} \label{fig:Unification}
\end{center}
\end{figure}

If we assume that there is not one vector lepton family but three (one for each Standard Model generation), then gauge coupling unification is
achieved quite precisely. Figure \ref{fig:Unification} shows two-loop gauge coupling unification predictions for $\alpha_s(M_Z)$. The
predictions are within $\approx 2\sigma$ of the experimental uncertainty (the uncertainty in our calculation is $\approx 0.004$), and are in
fact slightly closer to the observed $\alpha^{exp}_s(M_Z)=0.119\pm 0.002$ than in minimal SU(5) SUSY GUTs that predict $\alpha_s(M_Z)=0.130\pm
0.004$ (neglecting thresholds)\cite{4DSusyPredictions}.
However, the predicted unification scale is $M_G\sim 10^{11}$ GeV, so that avoiding dangerous proton decay rates is difficult in GUT completions
of this model. This model also results in a large contribution to the precision electroweak observable $S$. With three families we have
$S\approx \frac{1}{\pi}$, which is nearly 3-$\sigma$ from current best fits (see section \ref{subsec:pew}).

Adding a heavy SU(3) adjoint fermion (a ``gluino'') to the adjoint/vector lepton model recovers the low-energy spectrum of Split-SUSY
\cite{Arkani-Hamed:2004fb,Arkani-Hamed:2004yi,Giudice:2004tc}, and unification is achieved quite precisely with a gluino mass of $m\sim 8$ TeV
and a unification scale of $M_G\sim 10^{16}$ GeV.  As discussed in Sec \ref{subsec:pew}, this also results in a large contribution to the $S$
parameter.

\section{Experimental Prospects and Constraints}\label{sec:Experiment}
\begin{figure}
\begin{center}
\includegraphics[width=12cm]{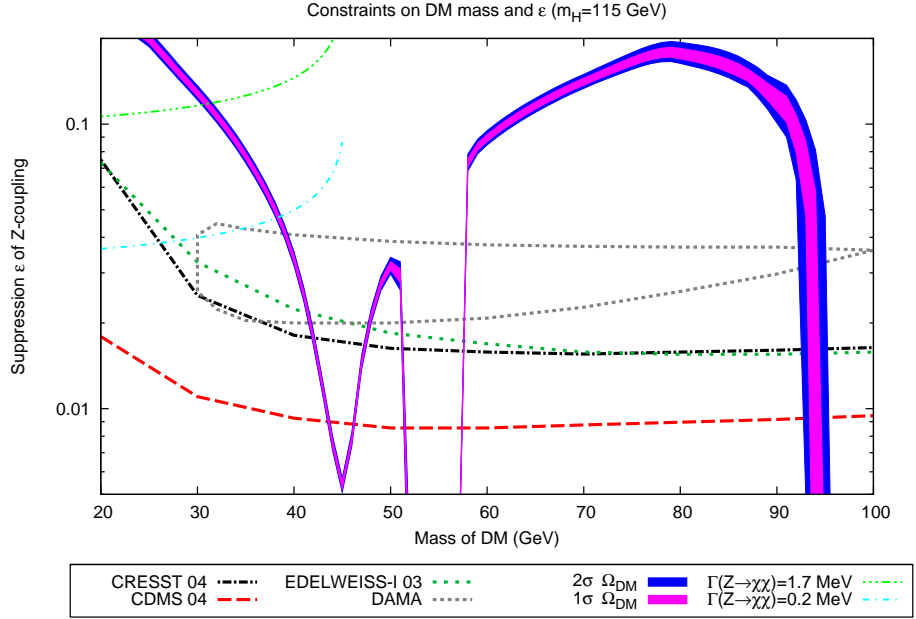}
\caption{Exclusion plot in the $\epsilon$--$m_{DM}$ plane for a {\bf Higgs mass of $m_H=115$ GeV}. The purple and blue bands correspond to 1-
and 2-$\sigma$ bounds on $\Omega_{nbm}=0.111 \pm 0.006$ from WMAP+CBI+ACBAR+2dFGRS data \cite{Eidelman:2004wy}. The blue and green curves on the
upper left correspond to the regions of partial $Z$ widths of 0.2 and 1.7 MeV, corresponding to 2- and 3-$\sigma$ exclusions from the LEP
$Z$-pole measurements, respectively \cite{Eidelman:2004wy}. The gray dotted contour is the 3-$\sigma$ region of the DAMA signal
\cite{Bernabei:2003sv}, while the remaining curves are the most recent published $90\%$ C.L. excluded regions from EDELWEISS
\cite{Sanglard:2005we}, CRESST \cite{Angloher:2005mb}, and CDMS \cite{Akerib:2004fq} (from top to bottom, respectively).}
\label{fig:Exclusion115}
\end{center}
\end{figure}

\begin{figure}
\begin{center}
\includegraphics[width=12cm]{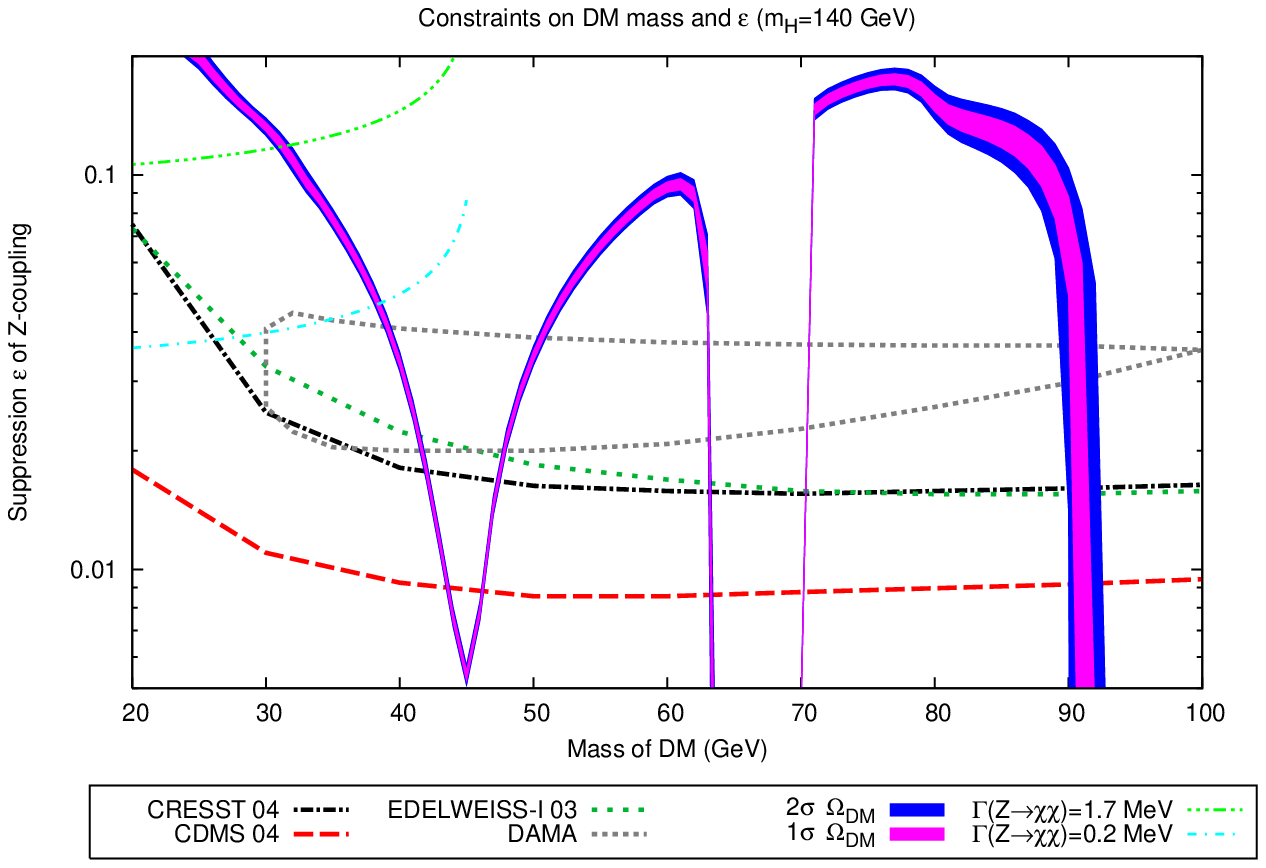}
\caption{The same as in Figure \ref{fig:Exclusion115} for a {\bf Higgs mass of $m_H=140$ GeV}}. \label{fig:Exclusion140}
\end{center}
\end{figure}

\begin{figure}
\begin{center}
\includegraphics[width=12cm]{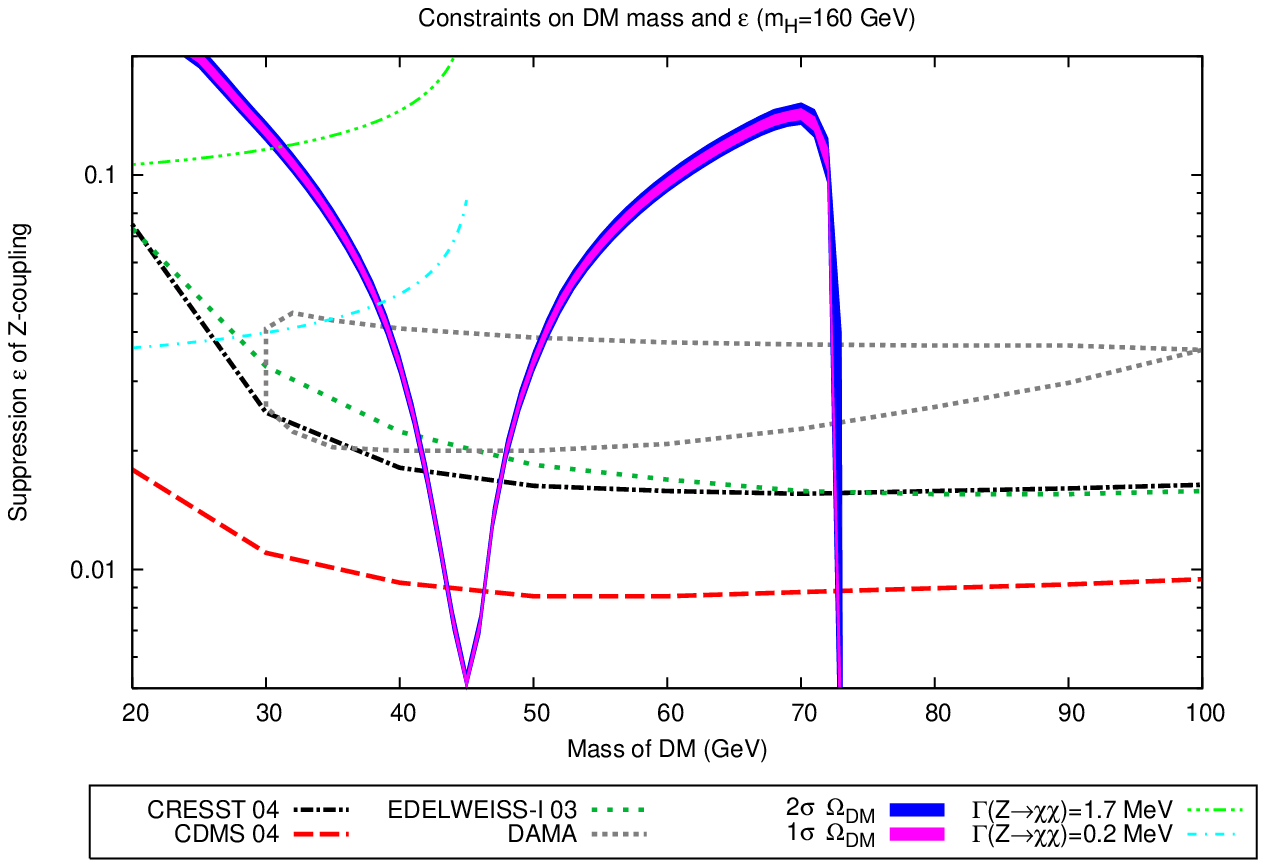}
\caption{The same as in Figure \ref{fig:Exclusion115} for a {\bf Higgs mass of $m_H=160$ GeV}} \label{fig:Exclusion160}
\end{center}
\end{figure}

In this section we discuss the implications of the dark matter scenario of the preceding sections for direct and indirect dark matter searches,
collider searches, and precision electroweak observables. Figures \ref{fig:Exclusion115}, \ref{fig:Exclusion140}, and \ref{fig:Exclusion160}
summarize experimental constraints on this model for $m_H=$ 115, 140, and 160 GeV. We display the $1\sigma$ and $2\sigma$ regions consistent
with $\Omega h^2=0.111\pm 0.006$ \cite{Eidelman:2004wy} (colored bands), the  regions of parameter space excluded by direct $Z$-pole
observations at LEP (upper left light green and blue curves are 3- and 2-$\sigma$ exclusions), and the constraints from direct dark matter
searches (processed as described below). The dashed gray contour is the boundary of the DAMA 95\% C.L. allowed region \cite{Bernabei:2003sv},
while the three lower curves represent the EDELWEISS \cite{Sanglard:2005we} (dotted dark green), CRESST \cite{Angloher:2005mb} (dot-dashed
black), and CDMS \cite{Akerib:2004fq} (dashed red) 90\% excluded regions.  Anything above the colored bands corresponds to a relic density below
the observed total abundance, i.e.\ a heavy neutrino comprising only a fraction of the dark matter.

The bounds from CDMS imply $\epsilon \lesssim 0.01$.  There are three mass ranges for which the dark matter is not overproduced in this regime:
at $m_{DM}\approx 45$, where even weakly coupled particles annihilate effectively through on-shell $Z$'s, at 90-95 GeV, and near the Higgs
resonance. In the latter two regions, the dark matter annihilates primarily through the Higgs channel.

\subsection{Direct Dark Matter Searches}\label{subsec:directdm}
Direct searches for dark matter look for elastic and inelastic collisions between WIMPs in the local dark matter halo and heavy nuclei in
detectors. Current searches are most sensitive to spin-independent interactions (which are generally mediated by scalar and vector nucleon
currents) and probe the cross sections of these interactions using very heavy nuclei such as Ge, I, W or Xe
\cite{Gondolo:1996qw,Gascon:2005xx,Goodman:1984dc}. Current and future experiments probing spin-independent and spin-dependent WIMP-nucleon
scattering include DAMA \cite{Bernabei:2003sv}, CDMS \cite{Akerib:2004fq,Mandic:2005ma}, EDELWEISS \cite{Sanglard:2005we}, CRESST
\cite{Majorovits:2004fa,Angloher:2005mb}, PICASSO \cite{Barnabe-Heider:2005pg}, SIMPLE \cite{Girard:2005pt}, ZEPLIN \cite{Cline:2003pi}, XENON
\cite{Aprile:2005mz}, NAIAD \cite{Alner:2005kt}, GENIUS \cite{Klapdor-Kleingrothaus:2002di,Klapdor-Kleingrothaus:2002um} and HDMS
\cite{Klapdor-Kleingrothaus:2002pg}.

WIMP-nucleon scattering results are usually quoted assuming isospin invariance $G_p\approx G_n$, but for $\chi$ the coupling to neutrons is
$G_n=\frac{\epsilon G_F}{2\sqrt{2}}$, while the proton coupling is greatly suppressed: $G_p=-(1-4\sin^2(\theta_w))G_n\approx -0.076 G_n$
\cite{Gondolo:1996qw}. We naively correct for this by reducing quoted sensitivities by $\left(\frac{(A-Z)-0.076 Z}{A}\right)^2=$ $0.28$, $0.30$,
and $0.32$ for Ge, I, and W detectors respectively \cite{Giuliani:2005my}. This reduction is taken into account in the exclusion plots
\ref{fig:Exclusion115} through \ref{fig:Exclusion160}, but the cross-sections in the text are as quoted from the original papers, without the
correction for isospin-dependent couplings.

Although to date there has been no conclusive discovery of a WIMP signal, the DAMA experiment (a Na/I detector) has reported a $6.3 \sigma$
annually modulating signal attributable to a WIMP with mass $m_{DM}\approx 50$ GeV and WIMP-nucleon cross section $\sigma\approx 7\times
10^{-6}$ pb \cite{Bernabei:2003sv,Bernabei:2005hj}.  A mixed heavy neutrino with $\epsilon\approx 0.02-0.04$ would have a cross-section
$\sigma_{\chi-n}\approx 10^{-5}$ pb consistent with the DAMA signal.  We see also that there are several masses ($m_{DM} \approx$ 40 or 50 GeV)
close to the reported DAMA signal mass for which this coupling would reproduce the observed $\Omega_{DM}$.

The DAMA signal seems excluded  at the $99.8\%$ C.L. by the Ge detectors EDELWEISS and CDMS, which use stronger background discrimination
techniques \cite{Eidelman:2004wy,Mandic:2005ma,Sanglard:2005we}. The bound on WIMP-nucleon cross-section obtained from the CDMS null signal with
standard estimates of the halo density and local halo profile is near $\sigma_{WIMP-n}\lesssim 4\times 10^{-7}$ pb at $m_{DM}=50$ Gev (assuming
isospin-invariance) \cite{Akerib:2004fq}. This is well below the cross-section of $\sigma_{\nu-n}\approx 2\times 10^{-2}$ pb for a Standard
Model heavy neutrino. Our model is consistent with this CDMS limit if $\epsilon \lesssim 0.01$ and with the EDELWEISS and CRESST limits if
$\epsilon \lesssim 0.02$.  If the one-loop radiative generation of $\epsilon$ is dominant, we would expect $0.01 \lesssim \epsilon \lesssim 0.1$
(see Sec. \ref{subsec:toyvector}), and $\sigma_{\chi-n}\approx 10^{-6}- 10^{-4}$ pb. Current experiments are probing this regime, and appear to
favor smaller $\epsilon$ and hence charged states that are very nearly degenerate. Two-loop effects from the Standard Model generate
$0.001\lesssim \epsilon\lesssim 0.01$ ($10^{-8}\lesssim \sigma_{\chi-n}\lesssim few\times 10^{-6}$ pb), and place a floor on the minimum
$\epsilon$ that can be obtained without resorting to fine-tuning. Current and future direct searches will increase sensitivities to
$\sigma_{WIMP-n}\sim 10^{-9}$ pb, and so will detect a signal or rule out the viable parameter space for models of this type.

\subsection{Indirect Searches}\label{subsec:indirectdm}
Indirect signals of dark matter annihilations in the Earth, Sun or Galactic halo can also be used to probe dark matter. The decay products of
annihilating WIMPs concentrated by gravity can give rise to sizable fluxes of neutrinos, gamma rays and cosmic rays above the background of
conventional astrophysical sources. We discuss here the general outlook for indirect searches in the electroweak dark matter scenario. We leave
a detailed study of the indirect signatures of heavy maximally mixed neutrino dark matter for a future work.

As a large body travels through the local dark matter halo, collisions between dark matter particles and nuclei in the body can dissipate enough
energy that the dark matter particles become gravitationally bound and sink to the center of the body. An equilibrium is eventually expected
between the rate of capture and of annihilations enhanced by the greater density of dark matter in the center of the body. Excess high-energy
neutrino fluxes are the most reliable indicator of WIMP annihilations in the Sun or Earth, and result in bounds on spin-dependent and
spin-independent WIMP-nucleon cross sections. Current and future neutrino telescopes carrying out such observations include Super-Kamiokande
\cite{Desai:2004pq}, AMANDA \cite{Ackermann:2005sb}, BAIKAL \cite{Aynutdinov:2005hx}, ANTARES \cite{Brunner:2003fy}, NESTOR
\cite{Grieder:2001ha} and IceCube \cite{Yoshida:2004rn}. To date, no statistically significant neutrino excesses from the Sun or Earth have been
detected .

In addition to neutrinos, gamma rays and high-energy cosmic rays from the galactic center also offer a strong indirect detection possibility
\cite{Gondolo:1999ef}. Current and future experiments designed to measure gamma and cosmic ray fluxes include EGRET \cite{Mukherjee:1997qw},
HEAT \cite{Barwick:1997ig}, PAMELA \cite{Adriani:2003in}, AMS \cite{Lamanna:2003qr}, and GLAST \cite{Morselli:2003xw}. An excess in gamma rays
above 1 GeV detected by EGRET \cite{Mukherjee:1997qw,Strong:2004de} and in positrons peaking at 8 GeV by HEAT \cite{Barwick:1997ig} may be
products of WIMP annihilations in the galactic halo \cite{deBoer:2004xt,Baltz:1998xv,Baltz:2001ir}. Moreover, a possible excess of microwave
emission from the galactic center observed in WMAP microwave data is consistent with synchrotron radiation from $e^+e^-$ pairs produced by WIMP
annihilation \cite{Finkbeiner:2004us}.

The uncertainties in the annihilation rates and predicted flux excess due to uncertainties in the halo profile of the Milky Way make it
difficult to reach any conclusion regarding the significance of these signals, but we consider briefly their consistency with the model
presented here. Previous authors have studied indirect signals coming from heavy stable 4th generation neutrinos
\cite{Fargion:1998xu,Fargion:1998rh,Golubkov:1997ht,Belotsky:2004st}, but most recent work relating the observed gamma and cosmic ray excesses
to WIMP annihilations have focused on the MSSM neutralino. The principal tension in explaining the HEAT and microwave signals from neutralino
dark matter arises because, being Majorana fermions, their annihilations into light Dirac fermion final states such as $e^+e^-$ are suppressed.
As such, large bost factors are required to explain the strength of the HEAT positron signal \cite{Baltz:1998xv,Baltz:2001ir}. Moreover, as most
of the electrons and positrons from neutralino annihilations are indirect products, neutralino models tend to produce softer positron spectra,
whereas the analysis of WMAP data in \cite{Finkbeiner:2004us} may prefer harder spectra.

Our dark matter candidate is somewhat similar to the low-mass Kaluza-Klein state considered in \cite{Hooper:2005fj}, in that both are Dirac
particles. In that case, the HEAT signal could be explained with less need for large boost factors than in the neutralino scenario, and we
expect the same to be true for our model.  We further expect that, if $m_{DM}\lesssim 80$ GeV, our dark matter candidate would produce a harder
spectrum of positrons than in the neutralino case, and perhaps be more consistent with the WMAP ``haze''.  Because our dark matter is Dirac,
however, there may also be tension with the null neutrino flux signal from AMANDA\cite{Hooper:2005fj}. Further work is necessary to compare the
predicted gamma and positron spectra in our models to the EGRET, HEAT, and microwave observations.

\subsection{Collider Detection}\label{subsec:collider}
The most direct constraint on the dark matter particle itself, the lightest neutrino $\chi$, comes from the $Z$-width measurement at LEP. The
measured partial width to invisible states is $499 \pm 1.5$ MeV, obtained by subtracting the visible partial widths from the total width and
assuming lepton universality. The standard model predicts a partial width of $167.29 \pm 0.07$ MeV to each species of neutrino, or $501.81 \pm
0.13$ MeV overall \cite{Eidelman:2004wy}. If $\chi$ is lighter than $m_Z/2$, then it will contribute a partial width,
\begin{equation}
\Gamma_{Z\rightarrow \chi\chi} = (167.29 \pm 0.07) \epsilon^2 \sqrt{1-\frac{4 m_{DM}^2}{m_Z^2}},
\end{equation}
suppressed because of the reduced $Z$-coupling and a threshold suppression due to $\chi$'s finite mass. Regions of exclusion at two- and
three-$\sigma$ ($\Gamma_Z(\chi \bar\chi)=$ 0.2 and 1.7 MeV, respectively) are shown on the left side of the exclusion plots
\ref{fig:Exclusion115}, \ref{fig:Exclusion140}, and \ref{fig:Exclusion160}. It should be noted that the prediction of $\Gamma_{inv}$ from the
Standard Model alone is already $1.9 \sigma$ above the measured width.

A further constraint on this framework comes from the production of $\chi \chi'$ pairs through an off-shell $Z$ at LEPII. Though the $\chi \bar
\chi Z$ and $\chi' \bar\chi' Z$ couplings are suppressed by $\epsilon$, the mixed couplings $\chi \bar \chi' Z$ are proportional to
$\sqrt{1-\epsilon^2}$, and hence essentially unsuppressed. The maximum center of mass energy reached at LEP II was $\approx 200$ GeV, so $\chi
\chi'$ pairs would be produced if $m_{\chi}+m_{\chi'} \lesssim 200$ GeV. The $\chi'$ would decay in the detector to a stable $\chi$ and two jets
or two lepton tracks via an intermediate $Z$. Thus, we believe that a two-jet or two-lepton track plus missing energy signal would have been
seen at LEP II if the above mass condition were met. We are not aware of constraints in the literature on this decay mode, and a more careful
analysis of the statistics is required to determine the precise exclusion range.  Because it depends on $m_\chi+m_{\chi'}$, this bound is only
constraining when combined with the assumption of high-scale perturbativity of the Yukawa couplings, which requires $m_{\chi'}\lesssim 150$ GeV.

Because the dark sector interacts only weakly, and most of the energy in $pp$ collisions at the LHC will be in the form of energetic gluons,
production at the LHC will be accompanied by considerable backgrounds. The favored masses $m_{DM}\lesssim 90$ GeV for the dark matter $\chi$ and
$\lesssim 200$ GeV for the heavier neutral and charged states make these models accessible to the LHC, but it is unclear how long it will take
to obtain sufficient statistics to learn anything concrete about a given model \cite{Gianotti:2005fm}.

\subsection{Precision Electroweak Observables}\label{subsec:pew}
Any extension of the Standard Model that couples to the $SU(2)_L\times U(1)_Y$ sector contributes to the gauge boson self-energies and hence
modifies predictions for the oblique correction parameters $S$, $T$, and $U$ \cite{Peskin:1991sw}. These contributions are particularly large
for extensions involving fermions that get mass only from electroweak symmetry breaking, which do not decouple from the Standard Model even when
$m_f \gg m_Z$.

The T parameter measures the amount of custodial SU(2) breaking that occurs, (i.e. deviations from $\rho=1$), and current constraints imply that
$\sum_i{\frac{1}{3}\Delta m^2_i} \lesssim (85)^2$ GeV$^2$ at $95\%$ CL, where $\Delta
m^2=m_1^2+m_2^2-\frac{4m_1^2m_2^2}{m_1^2-m_2^2}\ln{\frac{m_1}{m_2}}$ and the sum is over all isodoublets \cite{Eidelman:2004wy}. $SU(2)_L$
fermion doublets also generically contribute positively to the $S$ parameter. Current fits suggest $S=-0.13\pm 0.10, T=-0.17\pm 0.12, U=0.22\pm
0.13$ \cite{Eidelman:2004wy}.

The contribution of a single-family vector lepton model to $S$ is $S\approx \frac{1}{3\pi}$ which is sufficiently small to not be strongly at
odds with precision electroweak data. Moreover, under the assumption that the vector lepton sector is close to strong coupling near the GUT
scale ($M_G\approx 10^{13}$ GeV in these models), the spectrum is not generically split by enough to contribute largely to $T$. Only in cases
where there is a light dark matter candidate (such as when the $SU(2)_R$ symmetry of section \ref{subsec:toyvector} is approximately preserved)
do we expect a sizable positive contribution to $T$. In this case, the experimentally preferred value of $S$ is closer to $0$, and hence more
consistent with the positive contribution to $S$. Further analysis is needed to work out the detailed predictions for electroweak observables
with a vector lepton family, but this minimal model is not ruled out.

For the adjoint/vector lepton model, however, contributions to $S$ are much larger, $S\approx \frac{1}{\pi}$. As with the vector model, the
contributions to $T$ depend on the detailed form of the spectrum but do not have to be large. Based on the $S$ parameter alone, this minimal
model is ruled out at $\approx 4\sigma$ (if there is a large contribution to $T$, then the exclusion is at $\approx 3\sigma$). An invariant mass
term $m_T$ for the adjoint field can be added while maintaining a dark matter mass arising entirely from electroweak symmetry breaking so long
as $m_T\lesssim M_Z$, thereby suppressing the adjoint field's large contribution to $S$.

\section{The Cosmological Constant, Gauge Hierarchy, and the Structure Principle}\label{sec:structure}
In this section, we discuss some of the implications of the ``structure principle'' as defined in \cite{Arkani-Hamed:2005yv}. For our purposes,
this principle requires that large-scale structure develop in the universe as it cools. In the context of landscape scenarios, this principle is
justified by the weak anthropic argument that biological creatures will not develop in a universe that cannot support the development of stars
and other large-scale stable structure. We review how this principle was first applied by Weinberg to explain the smallness of the cosmological
constant and discuss an application suggested in \cite{Arkani-Hamed:2005yv} to a possible resolution to the gauge hierarchy problem in the
vector lepton model.

\subsection{Predicting the cosmological constant.}\label{subsec:lambda}
Weinberg's argument predicting a cosmological constant that is very small if nonzero begins with the observation that gravitational collapse via
Jeans instabilities can occur only after the universe becomes matter dominated. Moreover, linear sub-horizon perturbations to the energy density
can only grow when the energy density in the form of a cosmological constant is smaller than the energy density of matter, so the universe must
be in this regime when non-linear structures such as galaxies start to form. After matter-radiation equality, initial perturbations
$\frac{\delta \rho}{\rho}$ scale as the acceleration parameter $a$, so non-linear structures begin to form after the universe has expanded by an
amount $(\frac{\delta \rho}{\rho})^{-1}$ after matter-radiation equality, and we require for structure formation $\Lambda\lesssim
\rho_{MR}(\frac{\delta \rho}{\rho})^{3}$, where $\rho_{MR}$ is the energy density at matter-radiation equality.

As was pointed out in \cite{Arkani-Hamed:2005yv}, if the dark matter is dominated by cold relics a standard perturbative freeze-out calculation
gives
\begin{equation}
\rho_{MR}\approx \frac{1}{3g_{*}}\left(\frac{10^2}{M_{PL}\langle\sigma v\rangle}\right)^4 ,
\end{equation}
where $\langle\sigma v\rangle$ is the thermal average of the annihilation rate and $g_{*}$ is the number of effective
degrees of freedom. Weinberg's argument then bounds $\Lambda^{\frac{1}{4}}$ as,
\begin{equation}
\Lambda^{1/4}\lesssim \frac{(\frac{\delta \rho}{\rho})^{3/4}}{(3g_{*})^{1/4}}\frac{10^2}{M_{PL}\langle\sigma v\rangle}
.
\end{equation}
Without any additional reason for $\Lambda^{\frac{1}{4}}$ to be small, this bound should be roughly saturated, so it is a generic prediction of
the structure principle.

For heavy (i.e. more massive than $M_Z$) weakly interacting CDM particles, the annihilation cross section $\sigma v\approx
\frac{\alpha^2}{m_{DM}^2}$. The above argument then implies
\begin{equation}
\Lambda^{1/4}\lesssim \frac{(\frac{\delta \rho}{\rho})^{3/4}}{\alpha^2}\frac{m_{DM}^2}{M_{PL}} ,
\end{equation}
as pointed out in \cite{Arkani-Hamed:2005yv}. Empirically, heavy WIMP scenarios lead to $\Omega h^2\approx .1(\frac{m_{DM}}{TeV})^{-2}$, thereby
implying that $m_{DM}$ is of order a TeV and that $\Lambda^{1/4}\sim \frac{v^2}{M_{PL}}$.

Although weakly interacting TeV-scale dark matter seems to be empirically consistent, there is no \emph{a priori} reason why $m_{DM}$ is so much
smaller than $M_{PL}$ or so close to $v$. One simple possibility is that weak scale supersymmetry implies that $m_{DM}\approx v$ as would also
be the case for any other solution to the hierarchy problem that also contains a dark matter candidate. If we give up the assumption that the
solution to the gauge hierarchy problem also explains dark matter, a simpler possibility emerges: {\it If the dark matter particle gets mass
only via electroweak symmetry breaking, then $m_{DM}$ will naturally be close to $v$, and $\Lambda^{1/4}\sim \frac{v^2}{M_{PL}}$ is predicted!}

\subsection{Generating the gauge hierarchy with the structure principle}
\label{subsec:gaugehierarchy} In addition to providing an explanation for the smallness of the cosmological constant, the structure principle
also suggests a possible explanation of the gauge hierarchy in any model with a Standard Model Higgs and a dark matter candidate that gets mass
only via electro-weak symmetry breaking, such as the vector lepton model discussed here. As we will show, electroweak symmetry breaking via a
negative $m_H^2$ much below $M_{PL}$ now becomes essential for structure formation. If the Higgs mass parameter $m_H^2$ can be scanned in a
landscape scenario and electroweak symmetry breaking can only happen in a very small window about its current value, then we will have found a
possible explanation for the hierarchy.

To see how this might work, we consider a suggestion first made in \cite{Arkani-Hamed:2005yv}. In the SM with a large top Yukawa coupling to the
Higgs, the quartic coupling $\lambda$ has a negative UV fixed point $\lambda_{UV}\approx -y_t^2$ and an IR fixed point $\lambda_{IR}\approx
+y_t^2$. If the Higgs is sufficiently light (i.e. $\lambda$ is small), then for energies beyond some threshold $M_{cross}$, $\lambda$ will be
driven negative by RG running and the vacuum becomes meta-stable. If $|\lambda(\mu)|$ does not become too large in the UV, the vacuum in the
early universe has only a very small amplitude to decay as the universe cools, and the theory is safe from cosmological disaster. Turning this
picture around, suppose $\lambda(M_{*})$ starts negative at the cutoff scale, say $\lambda(M_{*})\approx \lambda_{UV}$. As the universe cools
$\lambda(\mu)$ crosses zero at a scale $M_{cross}$ exponentially suppressed with respect to the cutoff. Moreover, the low energy universe looks
very different depending on $m_H^2$.

Case I ($m_H^2\geq 0$): The higgs vacuum centered around zero is exactly stable and electro-weak symmetry is broken only by the QCD quark
condensate. Because $M_W\approx \alpha_2\Lambda_{QCD}$, sphaleron transitions operate all the way down to $\Lambda_{QCD}$, biasing zero baryon
number and washing out any baryon asymmetry in the universe down to the freeze-out level of $\frac{n_B}{s}\approx 10^{-19}$. (See
\cite{Arkani-Hamed:2005yv} for a discussion of this effect in more detail). The universe remains essentially radiation dominated and devoid of
baryons, thereby prohibiting structure formation. Thus, the structure principle would rule this range of parameters out.

Case II ($m_H^2\leq -M_{cross}^2$): The Higgs field is heavier than $M_{cross}$, so $\lambda(m_H)<0$ and the Higgs vacuum is unstable.
Fluctuations trigger $\langle \phi \rangle\rightarrow M_{PL}$. When the vacuum decays, the universe becomes dominated by a cosmological constant
$\Lambda\sim M_{PL}^4$, so no structure can form in this case either.

Case III ($-M_{cross}\leq m_H^2\leq 0$): Because electroweak symmetry is broken below the scale $M_{cross}$, all of the matter fields can obtain
sizable masses and the universe can become matter-dominated allowing structure to form.

Thus, if the dark matter gets mass from electroweak symmetry breaking, the prediction without any fine-tunings is that the Higgs mass should be
close to $M_{cross}$. To determine how close $|m_H^2|$ should be to $M_{cross}$ to avoid fine-tuning, we calculate the 1-loop effective
potential and look for the largest values of $|m_H^2|$ for which a meta-stable minimum develops in $V(\phi)_{1-loop}$ at nonzero $\phi$.
Following \cite{Arkani-Hamed:2005yv}, we approximate $\lambda(\mu)$ near $M_{cross}$ using an approximate solution to its RGE. Following
appendix \ref{app:VacuumStability}, we see that $\lambda(\mu)\approx -b\log(\frac{\mu}{M_{cross}})$ where $b$ is an RG coefficient and,
\begin{equation}
\frac{V(\phi)_{1-loop}}{M_{cross}^4}\approx
-\epsilon^2(\frac{\phi}{M_{cross}})-\frac{b}{2}\log(\frac{\phi}{M_{cross}})(\frac{\phi}{M_{cross}})^4,
\end{equation}
where $\epsilon=|m_H|/M_{cross}$.  For the Standard Model, $b\approx 0.076$ and numerical solutions indicate that a stable secondary minimum
develops for $\epsilon\leq 0.09$ at $\langle\phi\rangle\leq 0.4M_{cross}$. Typical values for slightly smaller $\epsilon$ are in the range of
$\langle\phi\rangle\leq 0.2M_{cross}$. Thus, without fine-tuning to a scale below that suggested by the structure principle, we would expect
physical Higgs masses in the range of $(0.1-0.2)M_{cross}$ for the Standard Model. With a more careful analysis of the one-loop effective
potential, we expect to find that the threshold mass is slightly higher than this.

In the Standard Model with a higgs mass of $115$ GeV, we calculated $M_{cross}$ for a top mass in the experimentally allowed range of
$169.2-188.5$ GeV. $M_{cross}\approx 7, 40, 120, 1350$ TeV for top masses of $m_t=188.5, 178.1, 174.3, 169.2$ GeV respectively. If the top were
$m_t\approx 188$ GeV, then a higgs mass of $115$ GeV would be entirely consistent with this scenario without fine-tuning. A top mass near its
central value of $m_t=178$ leaves a small hierarchy to deal with.

\begin{figure}
\begin{center}
\includegraphics[width=12cm]{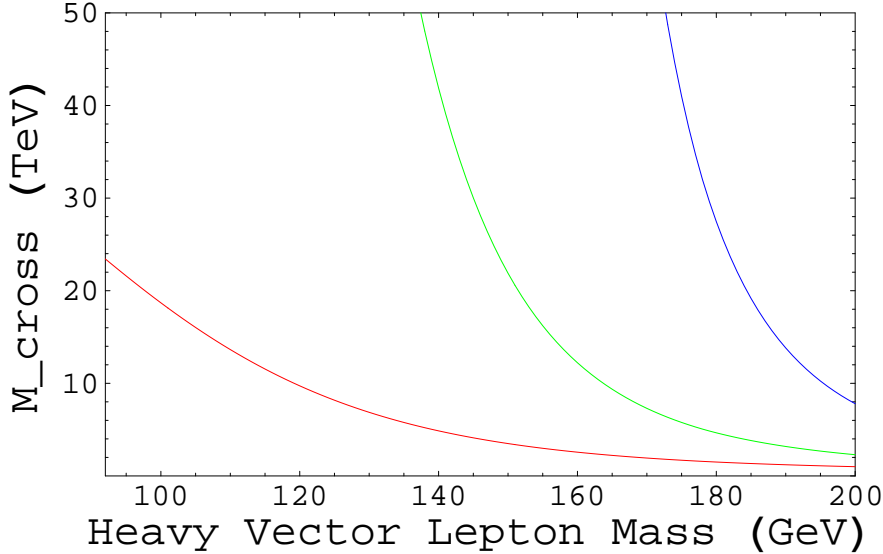}
\caption{$M_{cross}$ versus heavy degenerate charged and neutral state masses. The one-loop running of $\lambda$ was computed using tree level
values of the heavy lepton yukawa couplings. We used physical Higgs masses of $m_H=155, 135$ and $115$ GeV from top to bottom and a top mass of
$m_t=178.1$ GeV.} \label{fig:McrossNaive}
\end{center}
\end{figure}

Turning to the vector lepton model, the additional Yukawa couplings drive $\lambda(\mu)$ negative even faster. Thus, we expect to find
$M_{cross}$ lower for heavier lepton masses. Setting all heavy lepton masses equal to $M_V$, we consider the dependence on $M_V$ of $M_{cross}$
for several values of Higgs mass in figure \ref{fig:McrossNaive}.
The hierarchy between typical values of $M_{cross}$ in the range of 10-50 TeV in the Standard Model is eliminated for a light Higgs and a
modestly heavy vector lepton spectrum of $\sim 150$ GeV, but $\lambda(\mu)$ runs dangerously negative in the UV.

In the Standard Model, requiring that $\lambda(\mu)$ not run so negative that the vacuum should have decayed during
the last $10^{10}$ years leads to a bound on the Higgs mass of $m_H\geq 115$ GeV. For the vector model extension, even
with a light spectrum near $\sim 100$ GeV, $\lambda(\mu)$ tends to run too negative in the UV. In our analysis, we
calculated the scale at which $\lambda(\mu)$ becomes sufficiently negative for the vacuum in our universe to have
decayed already (assuming $m_H\geq 115$ GeV). We discuss the details of this requirement in appendix
\ref{app:VacuumStability}. For this rough analysis, we require that $\lambda(\mu)$ not be less than
$\lambda_{decay}\approx -0.13$.
\begin{figure}
\begin{center}
\includegraphics[width=12cm]{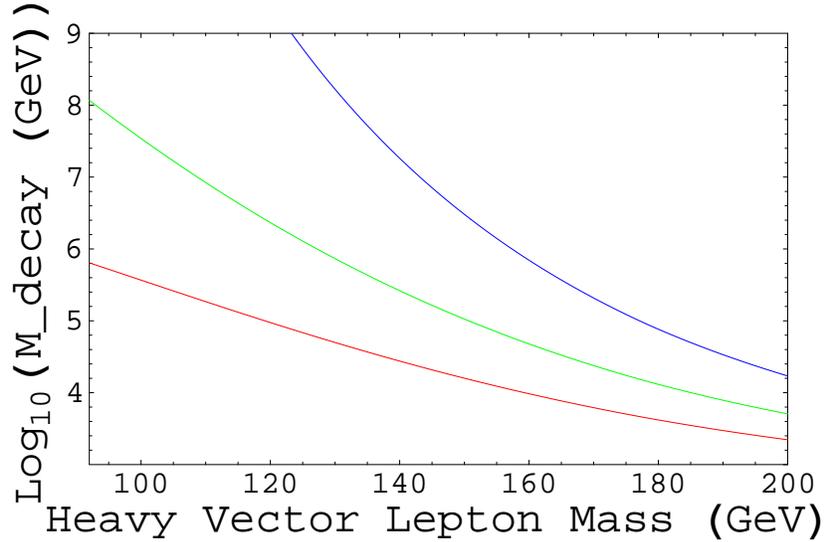}
\caption{The vacuum instability scale $M_{decay}$ versus heavy degenerate charged and neutral state masses. The one-loop running of $\lambda$
was computed using tree level values of the heavy lepton yukawa couplings. We used physical Higgs masses of $m_H=155, 135$ and $115$ GeV from
top to bottom and a top mass of $m_t=178.1$ GeV.} \label{fig:VacuumStabilityNaive}
\end{center}
\end{figure}
Figure \ref{fig:VacuumStabilityNaive} displays our results. We found that requiring $\lambda$ to not run past the stability bound requires
$m_H\gtrsim 155$ GeV. Thus, for the above explanation of the gauge hierarchy to work, new physics must enter below the scale $M_{decay}$ to
prevent $\lambda(\mu)$ from running too negative. This new physics should make the UV fixed point for $\lambda$ less negative. For example, new
$SU(3)$ fermions at an intermediate scale slows the running of $\alpha_3$ and hence helps keep the top Yukawa $y_t$ small, in turn making the UV
fixed point for $\lambda$ less negative. Alternatively, gauging the chiral symmetry that was introduced to forbid mass terms with a strongly
coupled U(1) that is broken near $M_{decay}$ (which requires adding an extra vector lepton generation to eliminate anomalies) drives the charged
and neutral state Yukawa couplings smaller in the UV, increasing the UV fixed point of $\lambda$.

Finally, the upper bound derived from the requirement of not hitting any Landau pole up to $M_G\approx 10^{13}$ GeV requires $m_H\lesssim 165$
GeV. The exact limit depends on vector lepton masses. Consequently, the self-consistency region for the Higgs mass in our minimal vector lepton
model is $155\lesssim m_H\lesssim 165$ GeV.

\section{Conclusions}\label{sec:conclusion}
We have considered a simple and well-motivated explanation for the origin of dark matter, namely that it consists of particles that get their
mass entirely through electroweak symmetry breaking. This framework connects the scale and existence of massive dark matter quite simply to the
electroweak scale. Moreover, this explicit connection affords new possibilities for explaining the smallness of the cosmological constant in
relation to the weak scale as well as the gauge hierarchy.

In the simplest models studied here, the dark matter candidate is a mixture of two Dirac neutrinos with opposite isospin, and so has a $Z$
coupling suppressed by $\epsilon=\cos(2\theta)$. Several approximate discrete and continuous symmetries can guarantee the radiative stability of
$\epsilon\approx 0$. If $\epsilon\lesssim 0.15$ is generated radiatively at one-loop order from breaking of the approximate symmetry within the
dark matter sector, the observed relic density can be accounted for with a dark matter mass in the range of $30-35$ GeV or $55-95$ GeV. This is,
however, disfavored by the null results of several direct WIMP searches.  The range $\epsilon\sim 0.01-0.001$ is typical if the approximate
symmetry is broken only by the Standard Model and $\epsilon$ is generated at two-loop order.  Then the dark matter mass is predicted to be
$m_{DM}\approx 45$ GeV or $m_{DM}\approx 90-95$ GeV and WIMP-neutron spin-independent cross sections are $\sigma_{WIMP-n}\sim 10^{-7}-10^{-8}$
pb. Current and future experiments sensitive to $\sigma_{WIMP-n}\sim 10^{-8}$ pb will probe this regime in the next few years and either detect
a signal or rule out the viable parameter space of this class of models.

We found that gauge couplings in a three-family vector lepton model unify quite well near $M_G\sim 10^{11}$ GeV. A
second model with the particle content of Split-SUSY unifies precisely at $M_G$ with a heavy ``gluino'' mass of $\sim 8$ TeV. Future work is
needed to build realistic ``top-down'' GUT models. Further work is also needed to examine consistency of the gamma ray and positron spectra
across the $\sim 1-100$ GeV range expected from WIMP annihilations in the galactic halo with HEAT and EGRET results, as well as to consider other
indirect search possibilities.

\section{Acknowledgements}
P.C.S. and N.T. would especially like to thank Nima Arkani-Hamed for inspiring this project and for many insightful discussions and comments.
Additional thanks to John Huth, Rakhi Mahbubani, Lubos Motl, Lisa Randall, Leonardo Senatore, and Christopher Stubbs for helpful comments
during the completion of this work. P.C.S. and N.T. are each supported by NDSEG Fellowships.

\appendix
\section{Renormalization Group Results For Vector Lepton Model}\label{app:rges}
\subsection{Vector Lepton Model RG Equations}
We have used the following $\beta$-functions in Sections 4 and 6. They are derived from the general expressions in
\cite{Machacek:1983tz,Machacek:1983fi,Machacek:1984zw}. Here $t=\ln \mu$, where $\mu$ is the renormalization scale.
 $g_1=\sqrt{\frac{5}{3}} g'$ is the coupling of the $SU(5)$-normalized $U(1)_Y$.  We quote two-loop beta functions
for the gauge couplings, one-loop for the Yukawa and quartic couplings.

\begin{enumerate}
\item Gauge couplings (2-loop)
\begin{align}
\betfn{g_i} = & g_i^3 b_i +\frac{g_i^3}{16 \pi^2}\biggl[ \sum_{j=1}^3 B_{ij} g_j^2 - \Tr\left(d^u_i \mathbf{y_u}^\dag
\mathbf{y_u}+d^d_i \mathbf{y_d}^\dag \mathbf{y_d}+d^e_i \mathbf{y_e}^\dag
\mathbf{y_e}\right)\nonumber \\
& -d^E_i \left(Y_1^2+Y_2^2\right) -d^S_i \tr(K^T K) \biggr]
\end{align}
with
\[ b=\barr{ccc}\frac{53}{10}&-\frac{5}{2}&-7\earr
\quad
B=\barr{ccc}\frac{28}{5}&3&\frac{44}{5} \\
\frac{27}{20}&14&12\\
\frac{11}{10}&\frac{9}{2}&-26\earr\]
\[d^u=\barr{ccc}\frac{17}{10}&\thh&2\earr
\quad d^d=\barr{ccc}\frac{1}{2}&\thh&2\earr\]

\[d^e=d^E=\barr{ccc}\frac{3}{2}&\frac{1}{2}&0\earr
\quad d^S=\barr{ccc}\frac{3}{10}&\frac{1}{2}&0\earr
\]

\item Yukawa couplings (1-loop):
\begin{eqnarray}
\betfn{\Yo{11}} &=& \Yo{11} \left(\Ys -\frac{9}{20} g_1^2-\frac{9}{4} g_2^2+\frac{3}{2} \Yo{11}^2
+\fr{3}{2}\Yo{12}^2+3 \Yo{21}^2-\fr{3}{2}
Y_1^2\right)+3 \Yo{12}\Yo{21}\Yo{22} \nonumber \\
\betfn{\Yo{12}} &=& \Yo{12}\left(\Ys -\frac{9}{20} g_1^2 - \frac{9}{4} g_2^2 + \frac{3}{2} \Yo{11}^2
+\fr{3}{2}\Yo{12}^2 + 3 \Yo{22}^2
-\fr{3}{2} Y_1^2 \right) + 3 \Yo{11}\Yo{21}\Yo{22} \nonumber \\
\betfn{\Yo{21}} &=& \Yo{21} \left(\Ys -\frac{9}{20} g_1^2 - \frac{9}{4} g_2^2 + 3 \Yo{11}^2 +\fr{3}{2}\Yo{21}^2 + \thh
\Yo{22}^2 -
\fr{3}{2} Y_2^2 \right) + 3 \Yo{12}\Yo{12}\Yo{22} \nonumber \\
\betfn{\Yo{22}} &=& \Yo{21} \left(\Ys -\frac{9}{20} g_1^2 - \frac{9}{4} g_2^2 + 3 \Yo{12}^2 +\fr{3}{2}\Yo{21}^2 + \thh
\Yo{22}^2 -
\fr{3}{2} Y_2^2 \right) + 3 \Yo{12}\Yo{12}\Yo{21} \nonumber \\
\betfn{Y_1} &=& Y_1\left( \Ys - \nf g_1^2 - \nf g_2^2 - \thh \Yo{11}^2 -
\thh \Yo{12}^2 + \thh Y_1^2 \right) \nonumber \\
\betfn{Y_2} &=& Y_2 \left( \Ys - \nf g_1^2 - \nf g_2^2 - \thh \Yo{21}^2 - \thh \Yo{22}^2 + \thh Y_2^2 \right) ,
\end{eqnarray}
with
\[\Ys=\Tr\left(3 \mathbf{y_u}^\dag \mathbf{y_u}+3 \mathbf{y_d}^\dag
\mathbf{y_d}+ \mathbf{y_e}^\dag \mathbf{y_e}\right) +Y_1^2+Y_2^2+\tr(K^T K)\label{ys}\] where $\Tr$ is over the three
Standard Model generations, and $\tr$ over the two dimensions of $K$.

\item Higgs quartic (1-loop)
\begin{align}
\betfn{\la}= &12 \la^2 -\left(\frac{9}{5} g_1^2+g_2^2\right) \la + \frac{9}{4}\left(\frac{3}{25} g_1^4+\frac{2}{5}
g_1^2 g_2^2+g_2^4\right)
\nonumber \\
&+4 \Ys \la -4 H(S),
\end{align}
with
\[H(S)=\Tr\left(3 (\mathbf{y_u}^\dag \mathbf{y_u})^2+
3 (\mathbf{y_d}^\dag \mathbf{y_d})^2+ (\mathbf{y_e}^\dag \mathbf{y_e})^2\right) +Y_1^4+Y_2^4+\tr((K^T K)^2)\]
\end{enumerate}

\subsection{Radiative Generation Of $\epsilon$}\label{app:radgen}
We note first that for small $\Delta$, the matrix
\begin{equation}\barr{cc} \kappa/\sqrt{2} (1+\Delta) &  \kappa'/\sqrt{2}(1+\Delta) \\
-\kappa/\sqrt{2}(1-\Delta) & \kappa/\sqrt{2} (1-\Delta) \earr
\end{equation}
can be diagonalized by a rotation of $\theta=\pi/4 + \frac{\kappa^2+\kappa'^2}{\kappa^2-\kappa'^2} \Delta$ on the left and $\frac{2 \kappa
\kappa'}{\kappa^2-\kappa'^2}\Delta$ on the right. The maximally mixed Yukawa matrix is of this form with $\Delta=0$, and all radiative breaking
effects we will consider can be parameterized as radiative generation of nonzero $\Delta$.  Thus, we can immediately relate a splitting $\Delta$
of the Yukawa matrix elements to the associated $Z$-coupling suppression $\epsilon = \cos(2 \theta)=2
\frac{\kappa^2+\kappa'^2}{\kappa^2-\kappa'^2} \Delta$.

For the estimates that follow, we work in the approximation that, although the eigenvalues $\kappa$ and $\kappa'$ of
$K$  also run, their fractional change is small enough that it does not greatly alter the running of $\Delta$.

We consider first the radiative effect of a splitting $Y_1=Y_2 (1+\delta)$ of the charged state Yukawa couplings at $M_G$, assuming maximally
mixed
$K$ at $M_G$. From the RGEs above,
\begin{equation}
\betfn{\Delta}=\frac{3}{4}(Y_2^2-Y_1^2) \approx \frac{3}{2} Y_2 \delta,\end{equation} so an order-of-magnitude estimate of $\epsilon$ at
the weak scale is given by
\begin{equation}
\epsilon(M_W) \sim \ln( \frac{M_G}{M_W}) \frac{1}{16 \pi^2} \frac{\kappa^2+\kappa'^2}{\kappa^2-\kappa'^2} 3 Y_2 \delta \sim \delta\end{equation}
assuming $\kappa'/\kappa \approx 1.5-2$.

In the case where $Y_1=Y_2$ at $M_G$ is exact, $\epsilon$ is not generated radiatively at one-loop, but there is a two-loop
contribution to $\beta(\Delta)$ from the Standard Model that goes as $y_t^2 g_2^2$. Naively, we expect
\begin{equation}
\epsilon_{SM}(M_Z) \sim \frac{1}{(4 \pi)^2} \ln( \frac{M_G}{M_Z}) \frac{\kappa^2+\kappa'^2}{\kappa^2-\kappa'^2} \alpha_t \alpha_2
N,\end{equation} where N accounts for color factors, the prefactor of the diagram in the RGEs, and the factor of 2 in the running of $\epsilon$.
We estimate a radiatively generated $\epsilon \sim 0.001$, though this could easily change by an order of magnitude if $N$ is large or small, or
because of the running of the parameters. Because this radiative generation appears to put $\epsilon$ in the region
of parameter space preferred to avoid direct detection, this radiative generation merits more careful study.

\section{Calculating Relic Dark Matter Abundances}\label{app:relicdens}
In this appendix, we review our calculation of the relic density for the heavy stable state $\chi$ of the vector
lepton model of section \ref{subsec:toyvector}. Our calculation is based on the discussion in \cite{Gondolo:1990dk},
\cite{Griest:1990kh}, and \cite{Edsjo:1997bg}.

Consider the evolution of the number density $n_1$ and $n_2$ for $\chi$ (with mass m)and its antiparticle
$\bar{\chi}$. Making the  standard assumptions of \cite{Gondolo:1990dk} and neglecting the co-annihilation
effects discussed in \cite{Edsjo:1997bg}, the evolution of the total number density $n=n_1+n_2$ is described by the
Boltzmann equation,
\begin{equation}
\dot{n}+3Hn=\frac{1}{2}\langle \sigma v_M\rangle (n^2-n^2_{eq}), \label{eq:Boltzmann}
\end{equation}
where $n_{eq}$ is the equilibrium number density, the M\/oller velocity $v_M$ is defined so that $v_Mn_1n_2$ is Lorentz
invariant, and $\langle \sigma v_M\rangle$ is the thermal average of the annihilation cross section calculated using
the numerical integral
\begin{equation}
\langle \sigma v_M\rangle = \frac{1}{8m^4TK_2^2(m/T)}\int_{4m^2}^{\infty}\sigma (s-4m^2)\sqrt{s}K_1(\sqrt{s}/T)ds
\label{eq:thermalaverage},
\end{equation}
where $K_i$ are modified Bessel functions of order $i$. It is useful to treat the effect of the expansion of
the universe implicitly by using the variable $Y=\frac{n}{s}$ where $s$ is the total entropy density of the universe.
We also use a dimensionless temperature variable $x=\frac{m}{T}$, so that \ref{eq:Boltzmann} becomes
\begin{equation}
\dot{Y}=-s \langle \sigma v_M\rangle (Y^2-Y_{eq}^2).
\end{equation}
From this we can obtain,
\begin{equation}
\frac{dY}{dx}=\frac{1}{3H}\frac{ds}{dx}\langle \sigma v_M\rangle (Y^2-Y_{eq}^2) ,
\end{equation}
where the Hubble parameter is $H=\sqrt{\frac{8}{3}\pi G\rho}$ with $G$ the gravitational constant and $\rho$ the
energy density of the universe. Using conventional definitions $\rho=g_{eff}(T)\frac{\pi^2}{30}T^4$, and
$s=h_{eff}(T)\frac{2\pi^2}{45}T^3$ \cite{Gondolo:1990dk}, we can finally re-cast the Boltzmann equation into the form
\begin{equation}
\frac{dY}{dx}=-(\frac{45G}{\pi})^{-1/2}\frac{g_*^{1/2}m}{x^2}\langle \sigma v_M\rangle (Y^2-Y_{eq}^2) ,
\end{equation}
where
\begin{equation}
g_*^{1/2}=\frac{h_{eff}}{g_{eff}^{1/2}}(1+\frac{1}{3}\frac{T}{h_{eff}}\frac{dh_{eff}}{dT}) .
\end{equation}

In our analysis, $\langle \sigma v_M\rangle$ was calculated numerically using equation \ref{eq:thermalaverage} and
then used to numerically solve the Boltzmann equation \ref{eq:Boltzmann}. Throughout, we assumed that the heavy charged
states all had mass $m_{heavy}=140$ GeV and the heavy neutral state $m_{\chi'}\approx 1.3 \approx$ consistent with a
GUT-scale strong coupling assumption and sufficient for coannihilations to be negligible\cite{Griest:1990kh,Edsjo:1997bg}.
The annihilation channels included in our analysis are listed in Table \ref{ChannelTable}. Feyncalc \cite{Mertig:1990an}
was used to simplify the trace and Lorentz algebra involved in calculating the Feynman diagrams. Mathematica version 5.1
was used for all numerical calculations.

\begin{table}
\begin{center}
\begin{tabular}{|l|c|c|c|}
\hline Final State & s-channel & t-channel & u-channel \\
\hline $f\bar{f}$ & $Z$, h & $\emptyset$ & $\emptyset$ \\
\hline  $Z$ $Z$ & h & $\chi$, $\chi^{\prime}$ & $\chi$, $\chi^{\prime}$ \\
\hline  $W^+W^-$ & h, $Z$ & $E_1$ & $E_2$ \\
\hline  $Z$ h & $Z$ & $\chi$ & $\chi$ \\
\hline  h h & h & $\chi$ & $\chi$ \\
\hline
\end{tabular}
\caption{List of $\chi\bar{\chi}\rightarrow$ Final State annihilation channels included in our relic density analysis. The
entries in the table are the intermediate states contributing to annihilation into a given final state for a given channel.}
\label{ChannelTable}
\end{center}
\end{table}

\section{Higgs Vacuum Stability}\label{app:VacuumStability}
In this appendix we briefly review our calculation of the Higgs vacuum stability bound in the vector lepton model of
section \ref{subsec:toyvector}. The physics of vacuum decay in relativistic quantum field theories is discussed
in \cite{Coleman:1975qj,Coleman:1977py} and an analysis applied to the Standard Model can be found in
\cite{Isidori:2001bm}.

With the addition of a single vector lepton family coupled to the Standard Model as in section \ref{subsec:toy}, the
Higgs quartic running is modified to include yukawa contributions from the heavy charged and neutral states,
\begin{equation}
(4\pi)^2\beta_{\lambda}=12\lambda^2+4(3y_t^2+2y_E^2+\kappa^2)\lambda-4(3y_t^4+2y_E^4+\kappa^4).
\end{equation}
For a small Higgs mass and hence small $\lambda(m_h)$, the large Yukawa couplings $y_t$ and $y_e$ will drive
$\lambda(\mu)$ negative for large $\mu$, thereby making the vacuum unstable. We wish to demand that in the lifetime of
the universe and neglecting finite-temperature effects, the probability for the vacuum to decay is less than 1. This 
requirement furnishes a lower bound on $\lambda$. Applying the results of \cite{Isidori:2001bm}, the decay rate per 
unit volume for the nucleation of a bubble of true vacuum of radius $R$ is $\frac{\Gamma}{V}\approx (\frac{1}{R})^4e^{-S_E}$, where
$S_E$ is the Euclidean action of the "bounce" solution (we are dropping multiplicative factors of order unity). Using
$S_E=\frac{16\pi^2}{3\lambda(1/R)}$, we take the probability for vacuum decay $p_d$ to be,
\begin{equation}
p_d\approx \frac{1}{H^4}\int{\frac{dR}{R} {\left(\frac{1}{R}\right)}^4e^{-\frac{16\pi^2}{3\lambda(1/R)}} },
\label{decayprobability}
\end{equation}
where $H=2.133h\times 10^{-42}=1.514\times 10^{-42}$ GeV ($h=0.71$) is the Hubble constant and $\frac{1}{H^4}$
accounts for the spatial and temporal extent of the universe. For the class of models under study, $\lambda$ runs
negative and then tracks its approximate UV fixed point. Thus, in practice the integral (\ref{decayprobability}) is dominated at
the scale where $\lambda$ is most negative, $M_{nuc}$ (this is typically the GUT scale $M_G\approx 10^{11}-10^{13}$
GeV in these models). In this case, we take, $p_d\approx M_G^4e^{-\frac{16\pi^2}{3\lambda(M_{G})}}$ to get
\begin{equation}
|\lambda(M_G)|\leq \frac{4\pi^2}{3\ln{\frac{M_G}{H}}} .
\end{equation}
Using $M_G\approx 10^{11}$ GeV, we find $\lambda_{decay}\approx -0.13$.

\end{document}